\documentclass[11pt]{article}
\usepackage[DIV=12]{typearea}
\usepackage{color}
\usepackage{amsmath,amssymb}
\usepackage{amsfonts}  
\usepackage{mathrsfs}
\usepackage{graphicx} 
\usepackage{subcaption}
\usepackage{bbm}
\usepackage{cite}
\usepackage{pifont}
\usepackage{enumerate} 
\usepackage{pdflscape}
\usepackage{hhline}  
\usepackage{multirow} %
\usepackage{booktabs} 
\usepackage{tabulary}
\usepackage[compat=1.1.0]{tikz-feynman}
\usepackage{nicefrac}
\usepackage{braket}
\usepackage{comment}
\usepackage{xspace}

\usepackage{cancel}
\usepackage{nicematrix}
\usepackage{bbold}
\usepackage{esvect}
\usepackage{arydshln}
\usepackage{tikz}
\usetikzlibrary{positioning}
\usepackage[normalem]{ulem}

\usepackage{xcolor}
\definecolor{darkgreen}{HTML}{109930}
\definecolor{pink}{rgb}{0.858, 0.188, 0.478}

\newcolumntype{R}[1]{>{\raggedleft\arraybackslash}p{#1}} 

\allowdisplaybreaks

\usepackage[pdftex,hidelinks,colorlinks=true,citecolor=darkgreen]{hyperref}
\hypersetup{
    pdftitle = {Conversions in multi-component dark sectors: a phase space level analysis},
    pdfauthor = {Chatterjee, Hryczuk}
}

\begin{document}

\date{}
\title{
  \vskip 2cm
  {\bf\huge Conversions in two-component dark sectors: a phase space level analysis}\\[0.8cm]
}
\author{
 {\bf\normalsize
  Shiuli Chatterjee\footnote{\texttt{shiuli.chatterjee@ncbj.gov.pl}},
  Andrzej Hryczuk\footnote{\texttt{andrzej.hryczuk@ncbj.gov.pl}} }\\[5mm]
 {\it\normalsize $^1$National Centre for Nuclear Research, Pasteura 7, 02-093 Warsaw, Poland}
}

\maketitle 

\thispagestyle{empty}

\vskip 1cm
\begin{abstract}
Conversions between the states in the dark sector affect not only their number densities but also their momentum distributions. In this work we study a phenomenologically motivated two-component dark matter scenario, based on the Coy Dark Matter model, in order to quantify the effect of conversions on departure from kinetic equilibrium and consequently the relic abundance. 
We perform a detailed numerical analysis at the level of the phase space distributions of dark sector particles, implementing all the relevant processes, including conversions, elastic scatterings and annihilations. Focusing on the parameter regions that lead to the observed relic abundance and provide a good fit to the Galactic Centre excess, we find that departure from kinetic equilibrium can alter the predictions for the total abundance by more than 100\%, while in most of the interesting parameter space being in the range from around -20\% to 50\%. The effect on each dark matter constituent separately can be much larger, even up to an order of magnitude, which can significantly affect the expected present-day gamma ray flux, and consequently phenomenology of the model.
\end{abstract}

\clearpage
\newpage
\section{Introduction}
The question of what constitutes about 25$\%$ of the energy density of the Universe continues to be a mystery, driving a dedicated theoretical study of dark matter (DM) and  efforts towards its detection. Weakly interacting massive particles (WIMP) are among the best motivated particle physics candidates for DM, favoured  for its simplicity, predictability and motivated connections to physics beyond the Standard Model (SM). However, a continued non-observance of any robust signatures  at colliders, at terrestrial direct detection experiments and from indirect detections has lead to progressively tighter constraints on the allowed parameter space for a WIMP DM (see e.g. \cite{Arcadi:2024ukq} for a recent review).

The standard treatment\cite{Gondolo:1990dk,Edsjo:1997bg} for calculation of DM relic densities  gives an efficient and highly accurate method of solving the integrated Boltzmann equation for the number density (nBE), relying on the assumption of kinetic equilibrium of DM (and co-annihilating partners, if any) with the Standard Model heat bath. 
However, the growing challenges to the WIMP paradigm have given cause to revisit this initial formulation and examine the validity of this assumption, finding that although the assumed kinetic equilibrium at freeze-out is \textit{typically} true for a canonical WIMP DM, an early departure from kinetic equilibrium can occur even in well-motivated, simplified models \cite{Aboubrahim:2023yag,Duch:2017nbe,Binder:2017rgn,Binder:2021bmg,Liu:2023kat,Du:2021jcj,Abe:2021jcz,Fitzpatrick:2020vba,Ala-Mattinen:2022nuj,Filimonova:2022pkj}, which then necessitates a solution of the full Boltzmann Equation (fBE) for the DM phase space distribution.
Global fit studies\cite{Chang:2022jgo,Chang:2023cki,GAMBIT:2021rlp} quantifying the extent of constraints on a given DM model from the latest experiments also benefit from such a treatment that provides higher precision in relic abundance calculations making it comparable to the current observational accuracy of percent level\cite{Planck:2018vyg}.

The challenges to WIMP paradigm have also spurred efforts for answers to the DM question to be sought in  an \textit{extended, non-minimal dark sector} (DS) allowing for, among other things, novel mechanisms of DM production\,\cite{Belanger:2011ww,DAgnolo:2017dbv,Garny:2017rxs,Brummer:2019inq,Kim:2019udq,Dror:2016rxc,Farina:2016llk,Puetter:2022ucx,Maity:2019hre}.
In beyond the Standard Model theories guided by symmetry principles, it isn't in fact uncommon to have an extended (dark) sector comprised of a multitude of particles transforming non-trivially under an additional symmetry, with the lightest particle being stable and hence the DM candidate. In a multicomponent DS, the number changing processes are multitudinous, including (inverse) decays, co-annihilations, inelastic scatterings and conversions, in addition to the canonical $2\rightarrow 2$ annihilations of DM to SM bath particles.
The usual argument of chemical decoupling preceding kinetic decoupling (KD) is then further challenged  given that
the elastic scatterings and the leading number changing processes need not even be related by any crossing symmetry (as is the case for canonical WIMP) and can be sourced from completely different interaction vertices. 
Together with the multiple number changing processes governed by different mass scales and having different rates, one can in general expect to generate quite non-equilibrium shapes of the phase space distributions of the DS particles.
It is however not easy to quantify a priori how these non-thermal shapes in the distribution functions carry over to the total relic abundances, and calls for a solution of the full Boltzmann equation at the phase space level 

Some recent studies carrying out a phase space level analysis for two-component DM, with some simplifying assumptions, have reported results with large or small non-equilibrium effects depending on whether or not the model considered had early kinetic decoupling and on how strongly the number changing processes depended on the shape of the distribution function. 
In ref.\,\cite{Beauchesne:2024zsq} and ref.\,\cite{Garny:2017rxs}, the fBE effect is found to be small from having not very momentum dependent number changing process and from restoration of kinetic equilibrium from inverse decays, respectively.
While a significant difference between fBE and nBE was reported for a two-component DM in ref.\,\cite{DAgnolo:2017dbv,Brummer:2019inq}  with the velocity dependence and early kinetic decoupling caused by subthreshold annihilations (and no decays/inverse-decays to enforce kinetic equilibrium). A study with the solution  of the Boltzmann equation for the first two moments for a two-component DS has been carried out in ref.\cite{Duan:2024urq} (see also \cite{Cervantes:2024ipg}).\\

In this work we aim to quantify the effect the conversion processes have on the departure from equilibrium and the evolution of phase space densities of DM in a two-component dark sector. We undertake this by studying an example extension of a DM model
of phenomenological interest: the pseudoscalar mediated Coy DM model\,\cite{Boehm:2014hva} which was introduced as a plausible DM explanation to an excess observed in the extended gamma ray flux from the Galactic Centre\,\cite{Hooper:2010mq}. It is worth noting that the distinguishing feature of this model, its suppressed elastic scatterings which is by construction to evade the otherwise inhibitive direct detection constraints, also leads to relatively early kinetic decoupling making the phase space level analysis necessary. We make an implementation of the full Boltzmann equation for a two-component dark sector by extending the publicly available code DRAKE \cite{Binder:2021bmg}.\footnote{This extension will be available in the next public DRAKE release.}
This allows us to explore how the region preferred by the GCE fit is affected by the solution from this improved analysis. 
As a by-product we also show how the multi-component extension of Coy DM allows for richer phenomenology (e.g. conversion-driven GCE explanation), irrespective of the non-equilibrium effects. \\

We start by discussing the double Coy DM that we consider in section\,\ref{sec:model}. In sec.\,\ref{sec:BM} we showcase an example demonstrating early kinetic decoupling and the non-thermal phase space distributions, motivating the undertaking of solving the full Boltzmann equation, the details for which we give in sec.\,\ref{sec:BE}. We present the results of our analysis in sec.\,\ref{sec:results} and conclude in sec.\,\ref{sec:conclusions}. 

\section{Doubled coy DM}
\label{sec:model}
The Coy DM model is phenomenologically motivated to explain the excess\footnote{See section 2.1 of ref.\,\cite{Leane:2022bfm} for a review of leading explanations for and the current status of the Galactic Centre excess.} in the observation of spatially extended $\gamma$-rays from the Galactic Centre with a spectrum that peaks at a few
GeV\cite{Hooper:2010mq}. 
A minimal WIMP DM that can potentially source this excess while also reproducing the observed relic abundance is ruled out by the  non-observation of scattering signatures at terrestrial direct detection experiments, necessarily arising from crossing-symmetry.
The Coy DM comprised of a singlet, fermionic DM and a pseudoscalar mediator, 
however, evades the otherwise prohibitive constraints from direct detection
by having an elastic scattering rate suppressed by the momentum transfer (which is small at these experiments) by virtue of a pseudoscalar mediated interaction between fermions\cite{Boehm:2014hva}.

We consider an extension of the model with  an additional fermion. The dark sector is then comprised of two Dirac fermions $\chi_1$ and $\chi_2$,  and a pseudoscalar $s$ mediating their  interactions with the SM. Assuming that the pseudoscalar has interactions with the SM fermions proportional to their Yukawa couplings, motivated by the  Minimal Flavour Violation (MFV) ansatz\cite{DAmbrosio:2002vsn}, the dark sector interaction Lagrangian is given as,
\begin{equation}
    \mathcal{L}\supset i\lambda_{\chi_1}(\bar{\chi}_1 \gamma_5\chi_1)s + i\lambda_{\chi_2}(\bar{\chi}_2 \gamma_5\chi_2)\,s + \sum_{f\in \text{SM}} y_f\lambda_y(i\bar{f}\gamma_5  f)\,s,
\label{eq:Lag}
\end{equation}
where couplings $\lambda_{\chi_1}, \lambda_{\chi_2}$, and $ \lambda_y$ are free parameters, $y_f\equiv \sqrt{2} m_f/v$ is the Yukawa coupling of SM fermion $f$ with $v\simeq 246$\,GeV the vacuum expectation value (vev) of the SM Higgs field. 

In the absence of any explicit mixing terms or any additional interactions, the Lagrangian in eq.\,\eqref{eq:Lag} leads to a 2-component DM model, with both $\chi_1$ and $\chi_2$ being stable. 
Even though the DS particles cannot convert from one to the other via decays, they can still convert via $2\leftrightarrow2$ processes $\chi_1,\chi_1\leftrightarrow \chi_2,\chi_2$. The tree level rate of these \textit{conversions} are 
independent of the coupling $\lambda_y$, while the rate of annihilations $(\chi_\texttt{i},\chi_\texttt{i}\leftrightarrow\textrm{SM},\textrm{SM})$ and elastic scatterings $(\chi_\texttt{i},\textrm{SM}\rightarrow\chi_\texttt{i},\textrm{SM})$ of the DS particles $(\texttt{\small{i}}=1,2)$ depends on $\lambda_\texttt{i}$ and $\lambda_y$.
For $M_s<M_{\chi_\texttt{i}}$ there exists also the number changing process $\chi_\texttt{i},\chi_\texttt{i} \rightarrow s,s$ (\texttt{\small{i}}=1,2), also independent of $\lambda_y$. 
The relative strengths of these various number changing processes, as well as their relative strength compared to the elastic scattering rates, together dictate the type of freeze-out mechanism.\footnote{In the original coy DM model, the coupling of pseudoscalar mediator was fixed to the Yukawa coupling with the only free parameter being its coupling to DM, since the only the product of these two couplings appears in the observables. However, with an additional DS particle now the relative strengths of conversions and annihilation for examples can change depending on the ratio of $\lambda_y/\lambda_{\chi_1}$ and  their product is no longer the only effective coupling of significance.} We therefore have 6 free parameters, the three couplings $(\lambda_{\chi_1}, \lambda_{\chi_2}, \lambda_y)$ and  $M_s, M_{\chi_1}, M_{\chi_2}$, the masses of the three additional particles $s$, $\chi_1$ and $\chi_2$, respectively. 
We also give here the expressions for the squared amplitudes, summed over \textit{both} initial and final state internal degrees of freedom, for the processes included in the following sections,
\begin{equation}
\label{eq:amps}
    \begin{split}        |\mathcal{M}|^2_{\chi_\texttt{i},\chi_\texttt{i}\rightarrow \textrm{SM},\textrm{SM}} &=4\lambda_{\chi_\texttt{i}}^2\lambda_y^2\,\frac{s^2}{(s-M_s^2)^2+\Gamma_s^2 M_s^2},\\
    |\mathcal{M}|^2_{\chi_1,\chi_1\leftrightarrow \chi_2,\chi_2} &=4\lambda_{\chi_1}^2\lambda_{\chi_2}^2\,\frac{s^2}{(s-M_s^2)^2+\Gamma_s^2 M_s^2},\\
    |\mathcal{M}|^2_{\chi_\texttt{i},\textrm{SM}\rightarrow \chi_\texttt{i},\textrm{SM}} &=4\lambda_{\chi_\texttt{i}}^2\lambda_y^2\,\frac{t^2}{(t-M_s^2)^2},\\
    |\mathcal{M}|^2_{\chi_\texttt{i},\chi_\texttt{i} \rightarrow s,s}&=2 \lambda_{\chi_\texttt{i}}^4\frac{\left(t-u)^2(M_{\chi_1}^4-M_s^4+t\,u-M_{\chi_1}^2(t+u)\right)}{(t-M_{\chi_1}^2)^2(u-M_{\chi_1}^2)^2},
    \end{split}
\end{equation}    
where $\Gamma_s$ is the total decay width of the pseudoscalar mediator $s$. There exist stringent constraints  on a light pseudoscalar coupling to SM, from flavour observables\,\cite{Dolan:2014ska}, though not relevant to the heavier pseudoscalars we focus our attention on. 
There exist also constraints on heavier pseudoscalars from  missing energy searches at LHC\cite{Boehm:2014hva,Buckley:2014fba} that are indicative of giving possibly relevant constraints with an updated study. This would require a dedicated study combining collider searches together with the details of the production of relic abundance studied here.  
Finally, there are analyses finding complementary constraints from Milky Way dwarf spheroidal galaxies (dSphs) favouring a muon-phillic explanation of the GCE over the one with MFV ansatz adopted in the coy DM scenario\cite{DiMauro:2021qcf,Abdughani:2021oit}.

\section{Benchmark -- an example to motivate fBE}
\label{sec:BM}
\begin{figure}
    \centering
    \begin{subfigure}{0.49\textwidth}
    \includegraphics[width=\textwidth]{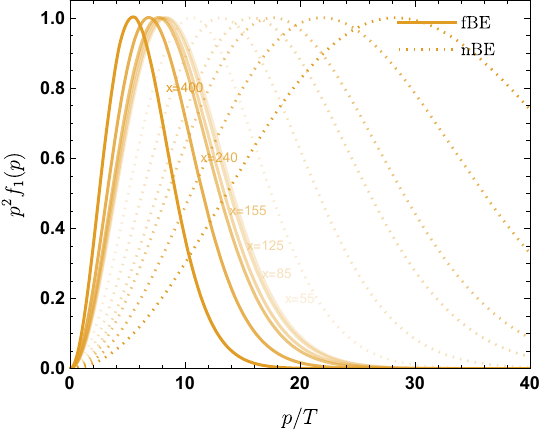}
    \end{subfigure}
    \begin{subfigure}{0.485\textwidth}
    \includegraphics[width=\textwidth]{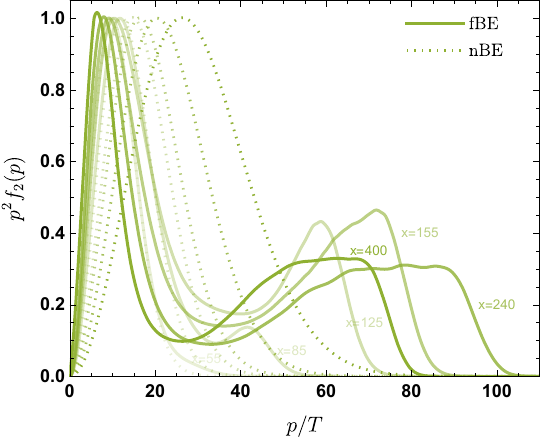}
    \end{subfigure}
\caption{Time snapshots of the evolution of the normalised momentum distributions for $\chi_1$ (left) and $\chi_2$ (right) in $p/T$, for the benchmark point showcasing the interplay of conversions and resonant annihilations with early kinetic decoupling. The solid lines show the particle distribution functions $f_{\chi_\texttt{i}}$ while the dotted lines show the corresponding equilibrium distributions.}
\label{fig:BM1}
\end{figure}

We showcase here an example point to demonstrate the impact of conversions, at the phase space level, in the thermal evolution of a multi-component dark sector. 
The following example point has the parameters:
$M_{\chi_1} = 44 $ GeV, 
$M_{\chi_2} = 38 $ GeV, 
$M_s = 80 $ GeV, 
$\lambda_1 = 0.0226 $, 
$\lambda_2 = 0.39 $, 
$\lambda_y = 0.3 $. 
Shown in fig.\,\ref{fig:BM1}  are the snapshots of the evolution of phase space distributions of $\chi_1$ and $\chi_2$. 
The distributions are normalised to 1 at each $x$, with the progressively dark colours corresponding to increasingly large values of $x$. The solid  coloured lines give the normalised distributions for the indicated $x$ values, as found from the solution of the phase space level Boltzmann equation, eq.\,(\ref{eq:beq_disc}).

The dashed lines in the figures show for comparison the (normalised) equilibrium distributions at corresponding values of $x$. The substantial difference between nBE and fBE solutions is a marker of early kinetic decoupling taking place for both DS particles. Moreover the $\chi_2$ phase space distribution goes through a noticeably non-trivial evolution, from an interplay of conversions and annihilation. 

We may understand the observed evolution by noting that this is a point where the annihilations and conversions are enhanced by a  resonance (which is quite broad, $\Gamma_s\sim0.15 M_s$) for initial momenta of about 12 GeV. Throughout most of the relevant time period this preferred momentum happens to lie in between the main bump from the peak of the thermal equilibrium shape inherited from the initial local thermal equilibrium (LTE), and the second bump arising from the conversions of $\chi_1$ to $\chi_2$, thus enabling the second bump from conversions to additionally fuel annihilations in the $x\sim50-100$ range. 
The observed departure from kinetic equilibrium will also have, as we will see later, a significant impact on the abundance of both the particles, and the predicted signal strengths in indirect detection.
It should be emphasized that such effects can only be captured by the solution of the Boltzmann equation for the phase space distributions, the explicit forms and details for which we give in the following section. 

\section{Boltzmann equations for phase space distributions in two-component dark sector} 
\label{sec:BE}
We assume that the pseudoscalar mediator $s$ remains in kinetic equilibrium with the SM plasma, since it  has a non-suppressed and thus efficient elastic scattering rate. This leaves  us with only having to solve for the phase space distributions of the two DS particles $\chi_1$ and $\chi_2$.
These are governed by the Boltzmann equation in the Friedman-Robertson-Walker spacetime as:
\begin{equation}
\label{eq:beq1}
  E_{\chi_\texttt{i}}\left(\partial_t-H p\partial_p\right)f_{\chi_\texttt{i}}=C_\text{ann.}[f_{\chi_\texttt{i}}] + C_\text{el.}[f_{\chi_\texttt{i}}]+C_{\chi_\texttt{i},\chi_\texttt{i}\leftrightarrow\chi_{\texttt{j}}\chi_{\texttt{j}}}[f_{\chi_\texttt{i}}] ,
\end{equation}
where, $f_{\chi_\texttt{i}}(t,p)$ are the distribution functions of the DS particles (with $\texttt{\small{i}}\in{1,2}$) depending on time $t$ and momentum $p$, $H$ is the Hubble expansion rate and the collision terms for the processes
of annihilations $(\chi_\texttt{i},\chi_\texttt{i}\leftrightarrow \textrm{SM},\textrm{SM}\,\&\,
\chi_\texttt{i},\chi_\texttt{i}\leftrightarrow s,s)$, elastic scatterings $(\chi_\texttt{i},\textrm{SM}\leftrightarrow \chi_\texttt{i},\textrm{SM})$ and conversions $(\chi_\texttt{i},\chi_\texttt{i}\leftrightarrow\chi_{\texttt{j}}\chi_{\texttt{j}})$ 
are given by $C_\text{ann.}, C_\text{el.}$ and $C_{\chi_\texttt{i},\chi_\texttt{i}\leftrightarrow\chi_{\texttt{j}}\chi_{\texttt{j}}}$, respectively. For the general structure of a collision term for a $2\leftrightarrow 2$ process\footnote{Note that had any of the DS states been unstable, one should also include in the collision term all the possible decay and inelastic scattering processes.}, see eq.\,$2$ of \cite{Hryczuk:2022gay}.  The collision term for conversions (with $\texttt{\small{i}}\neq \texttt{\small{j}}\in {1,2}$) convolutes the evolution of the two DS particles $\chi_1$ and $\chi_2$, requiring a  solution of the coupled system of equations.

As the Universe expands, the physical momentum redshifts as $p\rightarrow p/a$, where $a$ is the scale factor. We can get a dimensionless momentum-coordinate, factoring out this redshifting, by defining a  comoving momentum $q\sim  p/((g^s_\text{eff})^{1/3}T)$, where $T$ is the temperature of the SM plasma and $g^s_\text{eff}(T)$ is the effective number of entropy degrees-of-freedom (for which we use the results from \cite{Drees:2015exa}). 
The distribution functions long after kinetic decoupling are a constant in this co-moving momentum, making it a good momentum co-ordinate to solve for the evolution of the phase-space densities in. Choosing the heavier of the two DM masses to be the common mass scale, $m\equiv M_{\chi_1}$, we can define a dimensionless time-variable as $x\equiv m/T$.

We can carry out of a change of variables in eq.\,\eqref{eq:beq1}, and subsequently solve them  numerically by discretizing the set of coupled equations over $q$. 
It is worth mentioning at this point that in a two-particle coupled system the relevant range of physical momenta for the two particles can be significantly different, depending on whether one or both of them are in thermal/kinetic equilibrium, as also  on the kinematics. 
To accommodate this difference in momentum scales, we make use of two different functions to map from  the physical momentum to the comoving momentum, such that it is possible to work with a single, uniform grid in $q$, while also spanning over two different ranges in the physical momenta for $\chi_1$ and $\chi_2$. 
Even with a suitable choice of a functional form at an initial $x$, it is hard to choose optimally the momentum range such that it covers all the relevant scales throughout the entire evolution. For this reason, we solve the Boltzmann equation over intervals in $x$ with momentum range for each DS particle adapted anew for each interval. 
In practice it looks like this:
\begin{equation}
\label{eq:PQ}
    q(|\vec{p}|_{\chi_\texttt{i}},T)\equiv q_{A,\texttt{i}}+q_{B,\texttt{i}} \frac{|\vec{p}|_{\chi_\texttt{i}}}{(g^s_\text{eff})^{1/3}T},
\end{equation}
with $ q_{A,\texttt{i}}$ and $q_{B,\texttt{i}}$ being free parameters chosen to best cover the relevant range of momentum over the given interval in $x$ for each particle, $\texttt{\small{i}}\in {1,2}$. 
It is equally possible to have two completely different functional forms for the mapping for the two particles, but we find this choice of parameters to be the best for the case under study. Changing the variables as follows,
\begin{equation}
  \frac{\text{d}f_{\chi_\texttt{i}}(q,x)}{\text{d}t}
  =
  x\tilde{H}\frac{\partial f_{\chi_\texttt{i}}}{\partial x}-\frac{\partial f_{\chi_\texttt{i}}}{\partial q} \tilde{H}\left[p\left(1+\frac{1}{3}\frac{\text{d}(\log g^s_\text{eff})}{\text{d}(\log T)}\right)\left(\frac{\text{d}q}{\text{d}p}\right)_{\chi_\texttt{i}}+T\left(\frac{\text{d}q}{\text{d}T}\right)_{\chi_\texttt{i}}\right], 
\end{equation}
where $\tilde{H}\equiv H/[1+(1/3)\text{d}(\log g^s_\text{eff})/\text{d}(\log T)]\equiv H/(1+\tilde{g})$, we are lead to the Boltzmann equation in the form:
\begin{equation}
      \frac{ \partial f_{\chi_\texttt{i}}}{\partial x}=\frac{\partial f_{\chi_\texttt{i}}}{\partial q} \left[\frac{p}{x}(1+\tilde{g})\text{d}Q\text{d}P^{\chi_\texttt{i}}+\frac{m}{x^2}\text{d}Q\text{d}T^{\chi_\texttt{i}}\right] +\sum_{\text{proc.}}\mathcal{C}^{\,\text{proc.},\texttt{i}} .
\end{equation}
For a uniform grid in $q$ of length $N$, this  equation gives us a system of $2N$  coupled, ordinary first-order differential equations:  
\begin{equation}
       \frac{ \text{d} \vec{\mathbf{f}}^{\chi_\texttt{i}}}{\text{d} x}= \frac{\partial \vec{\mathbf{f}}^{\chi_\texttt{i}}}{\partial q} \left[\frac{\vv{PQ}^{\chi_\texttt{i}}}{x}\left(1+\tilde{g}\right)\vv{\text{d}Q\text{d}P}^{\chi_\texttt{i}}+\frac{m}{x^2}\vv{\text{d}Q\text{d}T}^{\chi_\texttt{i}}\right] +\sum_{\text{proc.}}\vec{\mathcal{C}}^{\,\text{proc.},\texttt{i}}, \quad \texttt{\small{i}}\in \small{\{1,2\}}.
\label{eq:BEQ2}
\end{equation}
The functions mapping from $q$ to $p^{\chi_\texttt{i}}$ and their derivatives, using eq.\,\eqref{eq:PQ}, are given as:
\begin{eqnarray}
    PQ^{\chi_\texttt{i}} &=& (g^s_\text{eff})^{1/3}\,T\,\frac{q(|\vec{p}|_{\chi_\texttt{i}},T)-q_{A,\texttt{i}}}{q_{B,\texttt{i}}},\\
    \text{d}Q\text{d}P^{\chi_\texttt{i}}
    &=&\frac{q_{B,\texttt{i}}}{(g^s_\text{eff})^{1/3}T}, \\
    \text{d}Q\text{d}T^{\chi_\texttt{i}}&=&-(1+\tilde{g})\frac{q(|\vec{p}|_{\chi_\texttt{i}},T)-q_{A,\texttt{i}}}{T}.
\end{eqnarray}
We have also defined rescaled collision terms $\mathcal{C}\equiv C/(E_{\chi_\texttt{i}} x \tilde{H})$, the structures for which we give in the following. \\

The collision terms for the three contributing processes: annihilations, elastic scatterings and conversions, have different structures, in particular in the dependence on the distribution functions that are solved for. It follows that different analytical simplifications can be done in each of them, and therefore we consider them separately. 

\subsection{Discretized collision term for annihilations}
The computation of a general collision term involves  multidimensional integrations over the phase spaces of all particles in the process excepting the particle whose evolution one solves for. For annihilations, the collision term can be simplified to a single angular integration and one remaining integration over energy under the assumption of CP invariance, and by neglecting the Bose enhancement and Pauli blocking factors for the usually massive final states with thereby  non-degenerate distributions $(f\ll1)$ \cite{Gondolo:1990dk}. 
This simplified collision term for annihilations  
$\chi_\texttt{i}(p_1),\chi_\texttt{i}(p_2)\rightarrow X(p_1)_3,\bar{X}(p_4)$
is given for each DS particle as\cite{Gondolo:1990dk,Binder:2021bmg}:
\begin{equation}
\begin{split}
     \mathcal{C}^\text{ann.}[f_{\chi_\texttt{i}}]
    &=\frac{g_{\chi_\texttt{i}}}{x \tilde{H}} \int \frac{\text{d}^3 p_2}{(2\pi)^3} v_\text{m\o l}\sigma_{\chi_\texttt{i},\chi_\texttt{i}\rightarrow X,\bar{X}} \left[f^{eq}(E^{\chi_\texttt{i}}_1)f^{eq}(E^{\chi_\texttt{i}}_2)-f^{\chi_\texttt{i}}(E^{\chi_\texttt{i}}_1)f^{\chi_\texttt{i}}(E^{\chi_\texttt{i}}_2)\right], \\
    &=\frac{g_{\chi_\texttt{i}}}{(2\pi^2) x \tilde{H}}
    \int \text{d}p_2 \langle\sigma v_\text{m\o l}\rangle_\theta(p_1^{\chi_\texttt{i}},p_2^{\chi_\texttt{i}})
    \left[f^{eq}(E^{\chi_\texttt{i}}_1) f^{eq}(E^{\chi_\texttt{i}}_2)
    -f^{\chi_\texttt{i}}(E^{\chi_\texttt{i}}_1) f^{\chi_\texttt{i}}(E^{\chi_\texttt{i}}_2)
    \right]     ,    
\end{split} 
\label{eq:Cann1}
\end{equation}
where $g_{\chi_\texttt{i}}$ is the spin degree of freedom of $\chi_\texttt{i}$, $X,\bar{X}$ stands for all the SM fermions and mediator $s$, $v_\text{m\o l}\equiv [s(s-4m_\texttt{i}^2)]^{1/2}/(s-2m_\texttt{i}^2)$ is the M\o ller velocity, and we define an angular averaged quantity $\langle\sigma v_\text{m\o l}\rangle_\theta \equiv \sum_X\frac{1}{2}\int \text{d}(\cos{\theta})\, v_\text{m\o l}\,\sigma_{\chi_\texttt{i},\chi_\texttt{i}\rightarrow X,\bar{X}}$. The equilibrium distribution for a given temperature $T$ is given by the Maxwell-Boltzmann distribution $f^{eq}(E)\equiv e^{-E/T}$, which is a good approximation to the full Bose-Einstein/Fermi-Dirac distribution in the non-relativistic  regime of freeze-out production.

After discretization over $q$, the integrations over $p_2$ in the above equation  can be rewritten as a sum over discretized momenta: 
$$\int \text{d}p_2 \equiv \sum_k \left(\frac{\partial p}{\partial q}\right)^{\chi_\texttt{i}}_k \frac{\Delta q}{2}\, W_k = 
\sum_k\frac{\Delta q}{2} \,\text{d}P\text{d}Q^{\chi_{\texttt{i}}}_k  \,W_k,$$
where $\Delta q$ is the width of the uniform grid in $q$, and  $W=\{1,2,2,\ldots,2,2,1\}$ is an array of length $N$ with the weights for trapezoidal integration. The remaining integration, which in general must be carried out numerically, is contained in the angular averaged quantity $\langle \sigma v_\text{m\o l}\rangle_\theta$. 
The distribution functions can also be seen to be vectors in this discretized momentum space:
\begin{equation}
\begin{split}
    f^{\chi_\texttt{i}}(E^{\chi_\texttt{i}})\rightarrow f^{\chi_i}(\vec{E}^{\chi_\texttt{i}})\equiv \vec{\mathbf{f}}^{\chi_\texttt{i}}, \\
f^{eq}(E^{\chi_\texttt{i}})\rightarrow f^{eq}(\vec{E}^{\chi_\texttt{i}})\equiv \vec{f}^{eq,\texttt{i}}.
\end{split}    
\end{equation}
The annihilation collision term can then be written as a matrix equation:
\begin{equation}
    \vec{C}^{\text{ann.},\chi_\texttt{i}}
    =\vec{f}^{eq,\chi_\texttt{i}}(\text{AM}_\texttt{i}\cdot \vec{f}^{eq,\chi_\texttt{i}}) 
    - \vec{\mathbf{f}}^{\chi_\texttt{i}}(\text{AM}_\texttt{i}\cdot \vec{\mathbf{f}}^{\chi_\texttt{i}}),
\end{equation}
where we define a  matrix in the discretized space: 
\begin{equation}
    (\text{AM}_\texttt{i})_{i,k}\equiv \frac{g_{\chi_\texttt{i}}}{(2\pi^2)x\tilde{H}}\text{d}P\text{d}Q^{\chi_\texttt{i}}_k \,\text{d}q \,W_k \,(p^{\chi_\texttt{i}}_k)^2 \langle\sigma v_\text{m\o l}\rangle_\theta(p^{\chi_\texttt{i}}_i,p^{\chi_\texttt{i}}_k) ,
\end{equation}
making explicit the dependence of $\langle \sigma v_\text{m\o l}\rangle_\theta$ on the relevant momenta 
and $\small{i,k}\in \{1,2,3,\ldots,N\}$ spans over the momentum grid. 
We define here  notation that matrix multiplications are signified  by ``$\cdot$" while the remaining products are to be understood as element-wise products, so $\vec{\mathbf{f}}^{\chi_\texttt{i}}(\text{AM}_\texttt{i}\cdot\vec{\mathbf{f}^{\chi_\texttt{i}}})\equiv \mathbf{f}^{\chi_\texttt{i}}_i (\text{AM}_\texttt{i}\cdot\mathbf{f}^{\chi_\texttt{i}})_i, \,i\in\{1,\ldots,N\}$ is a vector and there's no sum over $i$.

Given that we solve for the Boltzmann equations coupled in  $\chi_1$ and $\chi_2$, we find the full collision term for annihilations written in the basis of collated discretized distribution functions $\vec{\textbf{f}}\equiv\{\vec{\textbf{f}}^{\,\chi_1},\vec{\textbf{f}}^{\,\chi_2}\}$ and $f^{eq}\equiv\{\vec{f}^{eq,\chi_1},\vec{f}^{eq,\chi_2}\}$
as
\begin{equation}
\label{eq:Cann_disc}
\mathcal{\vec{C}}^{\,\text{ann.}}=\vec{f}^{\,eq}( \text{AM}\cdot \vec{f}^{\,eq}) -\vec{\textbf{f}}\,(\text{AM}\cdot \vec{\textbf{f}}) ,
\end{equation}
where, 
\begin{equation}
    \text{AM}\equiv 
    \left(\begin{array}{cc}
        1 & 0 \\
        0 & 0
    \end{array}\right)\otimes \text{AM}_\texttt{1} +
    \left(\begin{array}{cc}
        0 & 0 \\
        0 & 1
    \end{array}\right)\otimes \text{AM}_\texttt{2}.
\end{equation}
\\
Note that this matrix is independent of the unknown $f_{\chi_\texttt{i}}$ and contains the only remaining (angular) integration which is carried out numerically, making it the slowest part of the calculation of the annihilation collision term. Hence, we can pretabulate AM in $x$, for each $x$-interval of the evolution, and use interpolated values for a given $x$ to get a significant speed-up.

\subsection{Discretized collision term for elastic scatterings}
\label{sec:elsc}
In contrast to the annihilation collision terms, the collision terms for elastic scatterings and conversions contain distribution functions of the DS particles (that we solve for) in the final states, so that a reduction of the multidimensional integration similar to  that in eq.\,\eqref{eq:Cann1} is not possible\,\cite{Bringmann:2006mu,Binder:2021bmg}.
A possible strategy to proceed with the computation of the collision terms for elastic scatterings is to simplify the expression in the limit of small momentum transfer, to arrive at the Fokker Planck type operator\cite{Bringmann:2006mu,Binder:2016pnr}
\begin{equation}
\label{eq:FP}
    \mathcal{C}^{\,\text{el.}}[f_{\chi_\texttt{i}}]\simeq \frac{1}{2x}\frac{\gamma_X}{\tilde{H}}\left[T E^{\chi_\texttt{i}}\partial_p^2+\left(2T\frac{E^{\chi_\texttt{i}}}{p}+p+T\frac{p}{E^{\chi_\texttt{i}}}\right)\partial_p +3\right]f_{\chi_\texttt{i}},
\end{equation}
which is given after  discretization over $q$ as
\begin{multline}
\vec{\mathcal{C}}^{\,\text{el.},\chi_\texttt{i}}\simeq\frac{\gamma_X}{2x\tilde{H}}\Biggl[
    \frac{q_{B,\texttt{i}}^2 \vec{E}^{\chi_\texttt{i}}}{T\,(g^s_\text{eff})^{2/3}}\mathbb{D}_2+  \\
    \left(
    (\vec{q}-q_{A,\texttt{i}}\vec{\mathbf{1}})
    +T (\vec{q}-q_{A,\texttt{i}}\vec{\mathbf{1}})(\vec{E}^{\chi_\texttt{i}})^{-1} - 
\frac{2q_{B,\texttt{i}}^2 (g^s_\text{eff})^{2/3}}{T} \vec{E}^{\chi_\texttt{i}} (\vec{q}-q_{A,\texttt{i}}\vec{\mathbf{1}})^{-1} \right)\mathbb{D}_1 + 3\,\mathbb{1}     \Biggr]\cdot\vec{\mathbf{f}}^{\chi_\texttt{i}}    ,
\label{eq:FPdisc} 
\end{multline}
where $\vec{\mathbf{1}}=\{1,1,\ldots ,1,1\}$ is an array of length $N$ and we define inverses as $\left((E^{\chi_\texttt{i}})^{-1}\right)_i\equiv 1/E^{\chi_\texttt{i}}_i$ and $\left((\vec{q}-q_{A,\texttt{i}}\vec{\mathbf{1}})^{-1}\right)_i\equiv 1/(\vec{q}-q_{A,\texttt{i}}\vec{\mathbf{1}})_i$. The momentum transfer rate for   $\gamma_X(T)$ is defined as\,\cite{gondolo2012effect,kasahara2009neutralino,Binder:2016pnr,Binder:2021bmg}
\begin{equation}
    \gamma_{X}\equiv \frac{1}{3g_{\chi_\texttt{i}}M_{\chi_\texttt{i}}T}\int \frac{\text{d}^3k}{(2\pi)^3}g^\pm_X (\omega)[1\pm g^\pm_X(\omega)]\int _{-4k_{cm}^2}^0\text{d}t (-t)\frac{\text{d}\sigma}{\text{d}t}v,
\end{equation}
for scattering against $X\in \{\textrm{SM}, s\}$,
with $\omega\equiv \sqrt{k^2+m_X^2}$, $k_{cm}^2\equiv M_{\chi_\texttt{i}}^2k^2/(M_{\chi_\texttt{i}}^2+2\omega M_{\chi_\texttt{i}} +m_X^2)$, and the differential cross-section is $(\text{d}\sigma/\text{d}t)v\equiv |\mathcal{M}|^2_{\chi_\texttt{i},X\leftrightarrow\chi_\texttt{i},X}/ (64\pi k \omega M_{\chi_\texttt{i}}^2)$ with the scattering amplitude 
summed over \textit{both} initial and final state internal degrees of freedom and 
evaluated at $s\simeq M_{\chi_\texttt{i}}^2+2\omega M_{\chi_\texttt{i}} +m_X^2$.
The phase-space distribution of the scattering partner $X$ from the SM bath is given as usual by $g^\pm_X(\omega)\equiv 1/[\exp(\omega/T)\pm 1]$.

The discretization of momentum derivatives is implemented by central finite difference method of fourth order accuracy, using several neighbouring points. 
For the boundary condition at  numerical infinity, we introduce ghost points $f^{\chi_\texttt{i}}_{N+1}(x)=f^{\chi_\texttt{i}}_{N+2}(x)=0$ assuming vanishing phase-space density, while for the boundary at origin we implement instead a forward differentiation, precluding the need to introduce any ghost points.  
Matrices $\mathbb{D}_1$ and $\mathbb{D}_2$ are defined to encode this finite differentiation in a matrix notation. It is worth noting that the choice of a uniform grid in $q$ gives more accurate results for the derivatives evaluated by finite differentiation, compared to an arbitrary discretization in $q$.
Similar to the previous section, matrix multiplications are signified  by ``$\cdot$" and remaining products  are defined as an element-wise product, so for example $\left((\vec{q}-q_{A,\texttt{i}}\mathbf{1})\, \mathbb{D}_1\right)_{i,j}\equiv (\vec{q}-q_{A,\texttt{i}}\mathbf{1})_i\, (\mathbb{D}_1)_{i,j} $ is a matrix and there's no sum over $i$.

However, the Fokker-Planck form of the collision operator is an approximation which is very good for canonical WIMP but possibly less so for other DM models\cite{Hryczuk:2022gay}, see appendix~\ref{sec:app} for a discussion on the validity of the approximation. 
In the latter kind of scenarios where the expansion in small momentum transfer is no longer valid, one must resort to the full, unexpanded elastic scattering collision term. 
The  momentum integrals in the 
$2\leftrightarrow 2$ collision terms from eq.\,\ref{eq:beq1} can in general be reduced to a 4-dimensional integration, while 
it has been shown that for processes with squared amplitudes depending only on one Mandelstam variable, a reparametrization of the general collision term  makes it possible to reduce this further to a 2-dimensional integration that needs to be done numerically\,\cite{Aboubrahim:2023yag,Hannestad:2015tea,Hahn-Woernle:2009jyb}.
Noting that in our model, the squared-amplitudes for elastic scattering  are dependent only on Mandelstam $t$ while that for conversions are a function only of Mandelstam $s$, eqs.\,\eqref{eq:amps},  we follow ref.\,\cite{Aboubrahim:2023yag} to simplify the elastic scattering and conversion collision terms.  
In this formalism, the collision term for elastic scatterings 
$\chi_{\texttt{i}}(p_1),\textrm{SM}(p_2)\rightarrow\chi_{\texttt{i}}(p_3),\textrm{SM}(p_4)$ is
\begin{equation}
\label{eq:Cel1}
\begin{split}
\mathcal{C}^{\textrm{el.}}[f_{\chi_\texttt{i}}]&= 
\frac{1}{128\pi^3 g_{\chi_\texttt{i}} x\tilde{H}}\frac{1}{E_{\chi_\texttt{i}}|\vec{p}_1^{\,\chi_\texttt{i}}|} \int_{m_\texttt{i}}^\infty \text{d}E_3^{\chi_\texttt{i}}\int_{E_{2,min}^\textrm{SM}}^\infty \text{d}E_2^\textrm{SM} \,\Pi(E_1^{\chi_\texttt{i}},E_2^\textrm{SM},E_3^{\chi_\texttt{i}})\left[f_3^{\chi_\texttt{i}} f_4^{eq}-f_1^{\chi_\texttt{i}} f_2^{eq}\right], 
\end{split}
\end{equation}
where $E_{2,min}^\textrm{SM}=\text{Max}(m_\textrm{SM},E_3^{\chi_\texttt{i}}-E_1^{\chi_\texttt{i}}+m_\textrm{SM})$, $\alpha \equiv 1/(128\pi^3 g_{\chi_\texttt{i}} x\tilde{H})$, and the integration kernel $\Pi$ can be calculated analytically\footnote{For squared-amplitudes that are functions of more than one Mandelstam variable, the multidimensional integration reduces to a 2-dimensional angular integration which in general needs to be done numerically to get a counterpart of the integration kernel $\Pi(E_1^{\chi_\texttt{i}},E_2^\textrm{SM},E_3^{\chi_\texttt{i}})$, This adds to the computational cost, but the structure of the collision term thereafter follows similarly.} as:
\begin{equation}
    \Pi(E^{\chi_\texttt{i}}_1,E^\textrm{SM}_2,E^{\chi_\texttt{i}}_3) 
    \equiv \Theta(k_+-k_-)\int_{k_-}^{k_+}\text{d}k\, |\mathcal{M}_{\chi_\texttt{i},X\rightarrow\chi_\texttt{i},X}|^2.
\end{equation}
For the $t$-only dependent elastic scattering amplitude, we have:  
\begin{equation}
\begin{split}
  k&\equiv|\vec{k}|\equiv|\vec{p}^{\,\chi_\texttt{i}}_3-\vec{p}^{\,\chi_\texttt{i}}_1| 
  =|\vec{p}^{\,\textrm{SM}}_4-\vec{p}^{\,\textrm{SM}}_2| , \\
  k_+&\equiv \text{Min}(|\vec{p}^{\,\chi_\texttt{i}}_3|+|\vec{p}^{\,\chi_\texttt{i}}_1|,|\vec{p}^{\,\textrm{SM}}_4|+|\vec{p}^{\,\textrm{SM}}_2|), \\
  k_-&\equiv \text{Max}(\big||\vec{p}^{\,\chi_\texttt{i}}_3|-|\vec{p}^{\,\chi_\texttt{i}}_1|\big|,\big||\vec{p}^{\,\textrm{SM}}_4|-|\vec{p}^{\,\textrm{SM}}_2|\big|).
\end{split}    
\end{equation}
After discretization over $q$, the elastic scattering collision term has the form:
\begin{equation}
\begin{split}
\mathcal{C}^{\text{el.},\chi_\texttt{i}}_i
&=\alpha\,\frac{\Delta q}{2}\,\frac{1}{E^{\chi_\texttt{i}}_i|\vec{p}_i^{\,\chi_\texttt{i}}|}
\sum_{j=1}^N\frac{|\vec{p}^{\,\chi_\texttt{i}}|_j}{E_j^{\chi_\texttt{i}}}\text{d}P\text{d}Q^{\chi_\texttt{i}}_j \,W_j\, f^{eq}(E_i^{\chi_\texttt{i}}) \\
&\hspace{2cm}\times\int_{E_{2,min}^\textrm{SM}}^\infty \text{d}E_2^\textrm{SM} \,\Pi(E_i^{\chi_\texttt{i}},E_2^\textrm{SM},E_j^{\chi_\texttt{i}})f^{eq}(E_2^\textrm{SM}) 
\left(\frac{f_j^{\chi_\texttt{i}}}{f^{eq}(E_j^{\chi_\texttt{i}})}-\frac{f_i^{\chi_\texttt{i}}}{f^{eq}(E_i^{\chi_\texttt{i}})}\right), \\
&=\sum_{j=1}^{N}\mathcal{A}^{\chi_\texttt{i}}_{ij}\left(\frac{f^{\chi_\texttt{i}}_j}{f^{eq}(E^{\chi_\texttt{i}}_j)}-\frac{f_i^{\chi_\texttt{i}}}{f^{eq}(E_i^{\chi_\texttt{i}})}\right),
\end{split}
\end{equation}
where we have used detailed balance to rewrite the distribution functions from eq.\,\eqref{eq:Cel1} as:
$$\left(f_3^{\chi_\texttt{i}} f_4^{eq}-f_1^{\chi_\texttt{i}} f_2^{eq}\right)= f^{eq}(E_1^{\chi_\texttt{i}}) f^{eq}(E_2^\textrm{SM})\left(\frac{f_j^{\chi_\texttt{i}}}{f^{eq}(E_j^{\chi_\texttt{i}})}-\frac{f_1^{\chi_\texttt{i}}}{f^{eq}(E_1^{\chi_\texttt{i}})}\right)$$
to facilitate the definition of a matrix,
\begin{equation}
    \mathcal{A}^{\chi_\texttt{i}}_{ij}\equiv \alpha\,\frac{\Delta q}{2}\,\frac{1}{E^{\chi_\texttt{i}}_i|\vec{p}_i^{\,\chi_\texttt{i}}|}
\frac{|\vec{p}^{\,\chi_\texttt{i}}|_j}{E_j^{\chi_\texttt{i}}}\text{d}P\text{d}Q^{\chi_\texttt{i}}_j \,W_j\, f^{eq}(E_i^{\chi_\texttt{i}}) \int_{E_{2,min}^\textrm{SM}}^\infty \text{d}E_2^\textrm{SM} \,\Pi(E_i^{\chi_\texttt{i}},E_2^\textrm{SM},E_j^{\chi_\texttt{i}})f^{eq}(E_2^\textrm{SM}),
\end{equation}
which contains the only remaining integration over $E_2^{\textrm{SM}}$, which is done numerically and is the slowest part of the calculation of the elastic scattering collision term. We can arrive at a compact matrix equation for the collision term: 
\begin{equation}    \vec{\mathcal{C}}^{\,\text{el.},\chi_\texttt{i}}=\text{EM}^{\chi_\texttt{i}}\cdot \vec{\mathbf{f}}^{\chi_\texttt{i}},
\end{equation}
by defining
\begin{equation}
    \text{EM}^{\chi_\texttt{i}}\equiv \mathcal{A}\cdot\left(\mathbb{1} \,\,(\vec{f}^{eq,\chi_\texttt{i}})^{-1}\right)-(\mathcal{A}\cdot\vec{\mathbf{1}})\,\left(\mathbb{1} \,\,(\vec{f}^{eq,\chi_\texttt{i}})^{-1}\right),
\end{equation}
where again we define the inverse as $\left((f^{eq,\chi_\texttt{i}})^{-1}\right)_i\equiv 1/f^{eq,\chi_\texttt{i}}_i$, $\mathbb{1}$ is the identity matrix of size $N$, and $\vec{\mathbf{1}}=\{1,1,\ldots,1,1\}$ is an array 
of size $N$. 

Finally, collecting the elastic collision terms for $\chi_1$ and $\chi_2$ we can write the full collision term as:
\begin{equation}
\label{eq:Cel_disc}    \vec{\mathcal{C}}^\text{ el.}=\text{EM}\cdot \vec{\mathbf{f}},
\end{equation}
with
\begin{equation}
    \text{EM}=
    \left(\begin{array}{cc}
        1 & 0 \\
        0 & 0
    \end{array}\right)\otimes \text{EM}_\texttt{1} +
    \left(\begin{array}{cc}
        0 & 0 \\
        0 & 1
    \end{array}\right)\otimes \text{EM}_\texttt{2}.
\end{equation}
Note that we get one such matrix for each of the scattering partners $(X\in \{\textrm{SM},s\})$  so that $\text{EM}$ implicitly contains a sum over all $X$. In practice we pre-tabulate EM in $x$ for the particles with the largest momentum transfer rates $\gamma_X$, and use the Fokker-Planck collision terms from eq.\,\eqref{eq:FPdisc} for the remaining particles, striking a balance between efficiency and having accurate results when the Fokker-Planck approximation is not justified.

\subsection{Discretized collision term for conversions}
For the conversion process, the matrix structure is more involved. So for simplification of notation, we first discuss the form of the conversion collision term for $\chi_1$ evolution,
\begin{equation}
\label{eq:Cconv1}
\begin{split}    
\mathcal{C}_{\chi_1\chi_1\rightarrow \chi_2\chi_2}=\frac{1}{128\pi^3 g_{\chi_\texttt{i}} x\tilde{H}}\frac{1}{E^{\chi_1}_1|\vec{p}_1^{\,\chi_1}|} \int_{M_{\chi_2}}^{\infty} \text{d}E^{\chi_2}_3 \int_{E^{\chi_1}_{2,min}}^\infty \text{d}E^{\chi_1}_2\,\Pi(E^{\chi_1}_1,E^{\chi_1}_2,E^{\chi_2}_3)\\   \hspace{1cm}\times\left[f_{\chi_2}(E^{\chi_2}_3)f_{\chi_2}(E^{\chi_2}_4)-f_{\chi_1}(E^{\chi_1}_1)f_{\chi_1}(E^{\chi_1}_2)\right] ,
\end{split}
\end{equation}
where $E^{\chi_1}_{2,min}={\textrm{Max}(M_{\chi_1},E_3^{\chi_2}-E_1^{\chi_1}+M_{\chi_2})}$, and the integration kernel
$\Pi$ can be calculated analytically as:
\begin{equation}
    \Pi(E^{\chi_1}_1,E^{\chi_1}_2,E^{\chi_2}_3) \equiv \Theta(k_+-k_-)\int_{k_-}^{k_+}\text{d}k |\mathcal{M}_{\chi_1,\chi_1\rightarrow\chi_2,\chi_2}|^2,
\end{equation}
where, for the $s$-only dependent conversions,
\begin{equation}
\begin{split} 
  k&\equiv|\vec{k}|=|\vec{p}^{\,\chi_1}_1+\vec{p}^{\,\chi_1}_2|=|\vec{p}_3^{\,\chi_2}+\vec{p}_4^{\,\chi_2}|,\\
  k_+&\equiv \text{Min}(|\vec{p}^{\,\chi_1}_1|+|\vec{p}^{\,\chi_1}_2|,|\vec{p}^{\,\chi_2}_3|+|\vec{p}^{\,\chi_2}_4|), \\
  k_-&\equiv \text{Max}(\big||\vec{p}^{\,\chi_1}_1|-|\vec{p}^{\,\chi_1}_2|\big|,\big||\vec{p}^{\,\chi_2}_3|-|\vec{p}^{\,\chi_2}_4|\big|).
\end{split}    
\end{equation}
After discretization over $q$, the integrations over $E_3^{\chi_2}$ and $E_2^{\chi_1}$ in eq.\,\eqref{eq:Cconv1} can be rewritten as sums:
$$\int dE_3^{\chi_2} \equiv \sum_j \left(\frac{\text{d}E}{\text{d}p}\right)^{\chi_2}_j\left(\frac{\partial p}{\partial q}\right)^{\chi_2}_j \frac{\Delta q}{2}\, W_j = 
\sum_j \frac{|\vec{p}|_j^{\chi_2}}{E^{\chi_2}_j} \frac{\Delta q}{2}\, \text{d}P\text{d}Q^{\chi_2}_j  W_j,$$
$$\int dE_2^{\chi_1} \equiv \sum_k \left(\frac{\text{d}E}{\text{d}p}\right)^{\chi_1}_k\left(\frac{\partial p}{\partial q}\right)^{\chi_1}_k \frac{\Delta q}{2}\, W_k = 
\sum_k \frac{|\vec{p}|_k^{\chi_1}}{E^{\chi_1}_k} \frac{\Delta q}{2}\, \text{d}P\text{d}Q^{\chi_1}_k  W_k,$$
and the discretized conversion collision term for $\chi_1$ evolution has the form:
\begin{equation} 
    (\mathcal{C}_{\chi_1\chi_1\rightarrow \chi_2\chi_2})_i=\sum_{k,j=1}^{N} \mathcal{C}^{\,\textrm{conv.},\chi_1}_{ikj},
\end{equation}
with,
\begin{equation}
\begin{split}
    \mathcal{C}^{\chi_1\,\textrm{conv.}}_{ikj}&\equiv
    \frac{\alpha}{E^{\chi_1}_i|\vec{p}_{\chi_1}|_i}\left(\frac{\Delta q}{2}\right)^2
    \left(\frac{|\vec{p}_{\chi_1}|_k}{E^{\chi_1}_k}\text{d}P\text{d}Q_{k}^{\chi_1}W_k\right)    \left(\hat{\Theta}_{ikj}\hat{\Pi}_{ikj}\hat{\textbf{f}}^{\,\chi_2}_{ikj})\right)\left(\frac{|\vec{p}_{\chi_2}|_j}{E^{\chi_2}_j}\text{d}P\text{d}Q^{\chi_2}_jW_j\textbf{f}^{\,\chi_2}_j\right)    \\
    &\,\,\,\,- \frac{\alpha}{E^{\chi_1}_i|\vec{p}_{\chi_1}|_i}\left(\frac{\Delta q}{2}\right)^2
     \left(\frac{|\vec{p}_{\chi_1}|_k}{E^{\chi_1}_k}\text{d}P\text{d}Q_{k}^{\chi_1}W_k\textbf{f}^{\,\chi_1}_k\right)  \left(\hat{\Theta}_{ikj}\hat{\Pi}_{ikj}\textbf{f}^{\,\chi_1}_{i}\right)
     \left(\frac{|\vec{p}_{\chi_2}|_j}{E^{\chi_2}_j}\text{d}P\text{d}Q^{\chi_2}_jW_j\right),
\label{eq:conv_disc1}     
\end{split}
\end{equation}
with,
\begin{equation}
    \begin{split}    
    \hat{\Pi}^{\chi_1}_{ikj}&\equiv\Pi
    (E^{\chi_1}_i,E^{\chi_1}_k,E^{\chi_2}_j), \\
    \hat{\Theta}^{\chi_1}_{ikj}&\equiv\Theta(E^{\chi_1}_k-\textrm{Max}(M_{\chi_1},E^{\chi_2}_j-E^{\chi_1}_i+M_{\chi_2})) , \\
    \hat{\textbf{f}}^{\,\chi_2}_{ikj}&\equiv f_{\chi_2}(E^{\chi_1}_i+E^{\chi_1}_k-E^{\chi_2}_j),
    \end{split}
\end{equation}
where we linearly interpolate over $\textbf{f}$ when the combination $E_i^{\chi_1}+E_k^{\chi_1}-E_j^{\chi_2}$ has values off of the grid of discretized $E$.  
Note that $\mathcal{C}^{ \textrm{conv.},\chi_1}_{ikj}$ is an $N\times N\times N$ matrix capturing the conversions in the evolution equation of $\chi_1$. We can 
define vectors $\vec{V}_{\chi_1}$ with components  $(\vec{V}_{\chi_\texttt{i}})_i=(|\vec{p}_{\chi_\texttt{i}}|/E^{\chi_\texttt{i}})_i \text{d}P\text{d}Q^{\chi_\texttt{i}}_iW_i$ and define three-dimensional matrices 
\begin{equation}
\label{eq:conv_Achi}
(\hat{A}_{\chi_1})_{ikj}\equiv\hat{\Theta}^{\chi_1}_{ikj}\hat{\Pi}^{\chi_1}_{ikj}/\left(E_{\chi_1}|\vec{p}_{\chi_1}|\right)_i,     
\end{equation}
to rewrite eq.\,\eqref{eq:conv_disc1} as a matrix equation:  
\begin{equation}
    \mathcal{C}^{\,\text{conv.},\chi_1}_{i} = 
\alpha\Big[\left(\left(\hat{A}_{\chi_1}\,\hat{\textbf{f}}^{\,\chi_2}\right)\cdot\left(\vec{V}_{\chi_1}\, \vec{\textbf{f}}^{\,\chi_2}\right)\right)\cdot\vec{V}_{\chi_1}-\left(\left(\hat{A}_{\chi_1}\,\vec{\textbf{f}}^{\,\chi_1}\right)\cdot\vec{V}_{\chi_2}\right)\cdot\left(\vec{V}_{\chi_1}\, \vec{\textbf{f}}^{\,\chi_1}\right)\Big]_i.
\end{equation}
We can find an analogous expression for the collision term for conversions for $\chi_2$ evolution, with a corresponding $N\times N \times N$ matrix for  $\mathcal{C}^{\,\text{conv.},\chi_2}_{ikj}$.
The conversions for the two particles can be collected 
in the basis of the collated discretized distribution functions $\vec{\textbf{f}}\equiv\{\vec{\textbf{f}}^{\,\chi_1},\vec{\textbf{f}}^{\,\chi_2}\}$.
This single collision term for conversions written succinctly for $\chi_1$ and $\chi_2$ together is given as: 
\begin{equation}
\label{eq:Cconv_disc}
\left(\mathcal{C}_{\chi_1\chi_1\leftrightarrow \chi_2\chi_2}\right)_i= \alpha\Big[\left(\left(\hat{A}\, \hat{\textbf{f}}\right)\cdot\left(\vec{V}\, \vec{\textbf{f}}\right)\right)\cdot\vec{V}-\left(\left(\hat{A}\, \vec{\textbf{f}}\right)\cdot\vec{V}\right)\cdot\left(\vec{V}\, \vec{\textbf{f}}\right)\Big]_i,
\end{equation}
where $\vec{V}\equiv\{\vec{V}_{\chi_1},\vec{V_{\chi_2}}\}$ and
$\hat{\textbf{f}}_{ikj}=\begin{cases}
    f_{\chi_2}(E^{\chi_1}_i+E^{\chi_1}_k-E^{\chi_2}_j) 
    , \text{  for}\, 1\leq i\leq N, \\  
    f_{\chi_1}(E^{\chi_2}_i+E^{\chi_2}_k-E^{\chi_1}_j) 
    , \text{  for}\, N+1\leq i\leq 2N.
\end{cases}$\\ and the collated three-dimensional matrix has the following structure incorporating the correct dot-products for the conversion collision terms:
\begin{equation}
\hat{A}_i\equiv \begin{cases}
    \left(\begin{array}{cc}
    0 & 1 \\
    0 & 0
\end{array}\right)\otimes (\hat{A}_{\chi_1})_i , \text{ 
  for, } 1\leq i\leq N, \\
  \left(\begin{array}{cc}
    0 & 0 \\
    1 & 0
\end{array}\right)\otimes (\hat{A}_{\chi_1})_{i-N} , \text{ 
  for,  } N+1\leq i \leq  2 N,
\end{cases} 
\end{equation}
written more explicitly as,
\begin{equation}
\hat{A}=\left\{
    \left(
    \setlength\arraycolsep{7pt}
    \def\arraystretch{1.3}
    \begin{array}{c|c}
       0  & \text{\footnotesize $(\hat{A}_{\chi_1})_{1}$} \\
       \hline 
       0 & 0
    \end{array}\right) ,
    \left(
    \setlength\arraycolsep{7pt}
    \def\arraystretch{1.3}
    \begin{array}{c|c}
       0  & \text{\footnotesize $(\hat{A}_{\chi_1})_{2}$} \\
       \hline 
       0 & 0
    \end{array}\right) ,
    \ldots ,
    \left(
    \setlength\arraycolsep{7pt}
    \def\arraystretch{1.3}
    \begin{array}{c|c}
       0  & 0 \\
       \hline 
       \text{\footnotesize $(\hat{A}_{\chi_2})_{1}$} & 0
    \end{array}\right) ,
    \left(
    \setlength\arraycolsep{7pt}
    \def\arraystretch{1.3}
    \begin{array}{c|c}
    \centering
       0  & 0 \\
       \hline 
       \text{\footnotesize $(\hat{A}_{\chi_2})_{2}$} & 0
    \end{array}\right) ,
    \ldots
    \right\},
\label{eq:conv_mat}
\end{equation}
using $\hat{A}_{\chi_\texttt{i}}$ as defined in eq.\,\eqref{eq:conv_Achi}.
The full conversion collision matrix $\hat{A}$ is of size $(2N)^3$, making it much larger than its counterparts in annihilations and elastic scatterings, and more CPU intensive. It gets this form because we need to  convert all the integrations in the collision terms to sums owing to the presence of the unknown distribution functions in the integrands. But this also means that there are no remnant numerical integrations to be evaluated. It is therefore not efficient to pre-tabulate the matrices over $x$, as we have done for annihilations and elastic scatterings, and we evaluate it at each step of the solver in $x$.  

We might add here that there also exist self-scattering processes for $\chi_1$ and $\chi_2$  mediated by $s$, which we 
have not included in eq.\,\eqref{eq:beq_disc}  since the self-scattering rates turn out to have a smaller impact than the remaining processes for most of the parameter space we study, though they may add some sub-leading corrections \cite{Hryczuk:2022gay}. Although in principle the self-scattering collision terms can be included with a structure analogous to the conversion collision terms, each such process requires the computation of large 3-dimensional matrices (of the kind in eq.\eqref{eq:conv_mat} of size $(2N)^3$) making the solution more CPU intensive. We leave this to future work.

\subsection{Discretized full Boltzmann equation} 
Finally substituting from  eqs.\,\eqref{eq:Cann_disc}, \eqref{eq:Cel_disc} and \eqref{eq:Cconv_disc} into the discretized Boltzmann equation eq.\,\eqref{eq:BEQ2}, we get the full Boltzmann equation discretized over $q$ in the form:
\begin{equation}
\begin{split}
    \frac{\text{d} \vec{\mathbf{f}}}{\text{d}x}&=\text{FM}\cdot \vec{\mathbf{f}} 
    +\text{EM}\cdot \vec{\mathbf{f}} +\vec{f}^{\,eq}(\text{AM}\cdot\vec{f}^{\,eq})
    - \vec{\mathbf{f}}\,(\text{AM}\cdot\vec{\mathbf{f}})  \\
    &\;\;\;\;\;\; + \alpha\Big[\left(\left(\hat{A}\, \hat{\textbf{f}}\right)\cdot\left(\vec{V}\, 
    \vec{\textbf{f}}\,\right)\right)\cdot\vec{V}-\left(\left(\hat{A}\, \vec{\textbf{f}}\,\right)\cdot\vec{V}\right)\cdot\left(\vec{V}\, \vec{\textbf{f}}\,\right)\Big],
\end{split}
\label{eq:beq_disc}
\end{equation}
where FM is the discretized form of the first term in eq.\,\eqref{eq:BEQ2} from the change of variables $(p,t)\rightarrow (q,x)$. The derivatives in momentum space are implemented by finite differentiation
method with fourth order of accuracy as described in section\,\ref{sec:elsc}. Adopted numerical approach relies on as much matrix formulation as allowed by the non-linear structure of the equation. For finding a solution we use implicit Adams-Moulton method of order two for time discretization with an intermediate time step for error control. The number of momentum grid points $N$ varies between 40 and 100 and was adjusted automatically at every integration step in $x$.

\section{Results}\label{sec:results}

We carry out a phase space level study of a multicomponent DS by way of example by splaying out this parameter space of the two-component coy DM with 6 free parameters.
The extension of the Coy DM model by an additional particle is an enlargement of the parameter space, so that the initial \textit{raison d'être} of the model, to explain the GCE while being commensurate with other DM observables, is still addressed.
However, the primary goal of our work is to study conversions at the phase space level, so we divide the full 6-parameter space into segments of interest where the observed relic abundance is reproduced by the sum of the abundances of the two DM particles, and the excess in the extended gamma ray signals at the Galactic Centre is explained by the sum of predicted fluxes from the two DM particles. 
We then narrow down upon parts of the parameter space where the conversion process plays a significant role in the freeze-out,
and carry out a phase-space level analysis of the evolution of DS particles, highlighting the impact on relic abundance and/or the extended gamma ray flux, 
thereby illustrating the importance of taking into account the phase space level evolution.

\subsection{Exploratory scans around the GCE region}

\begin{figure}[h!]
     \centering
     \begin{subfigure}[t]{0.512\textwidth}
         \centering         \includegraphics[width=\textwidth]{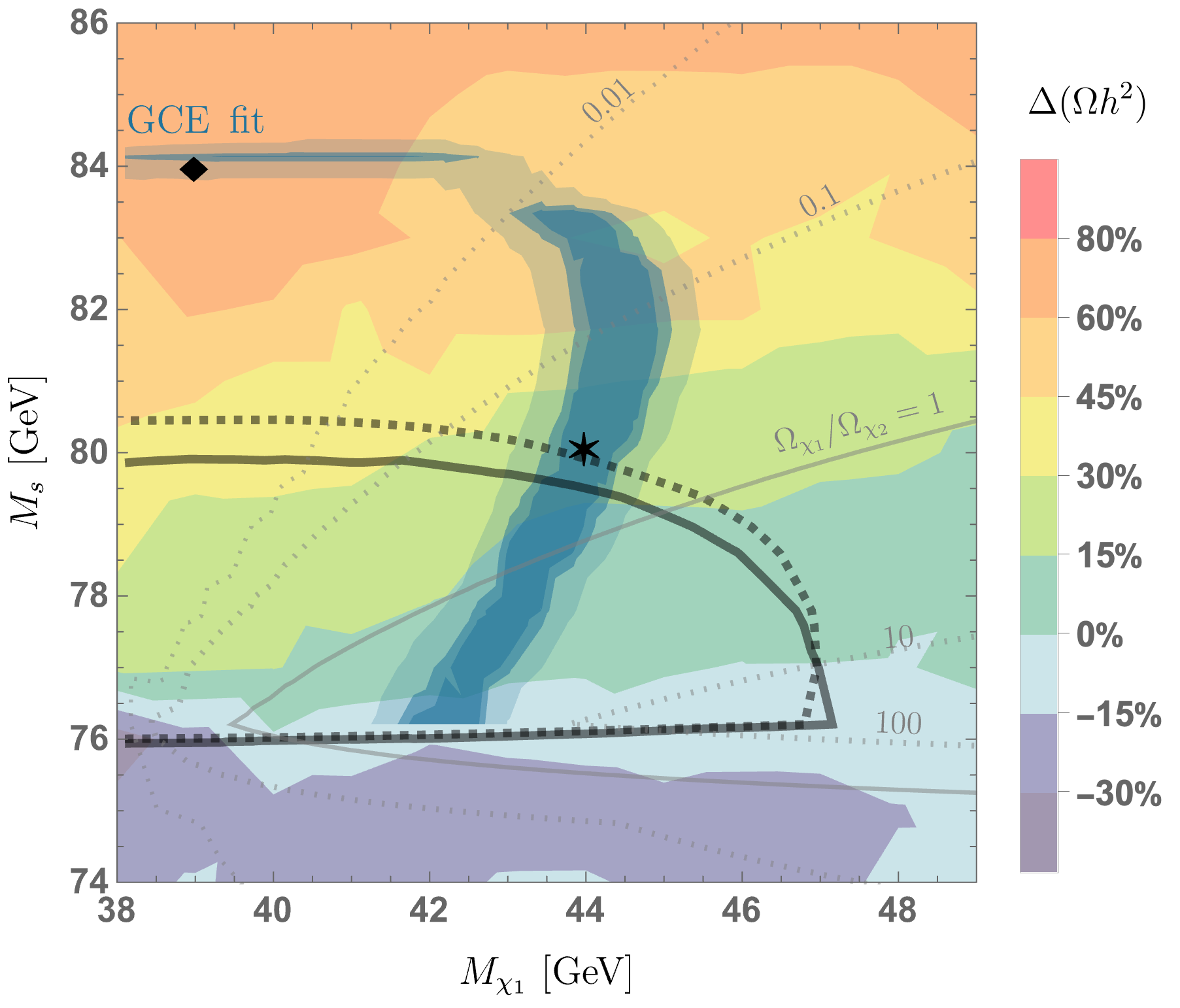}
     \end{subfigure}
     \begin{subfigure}[t]{0.478\textwidth}
         \centering         \includegraphics[width=\textwidth]{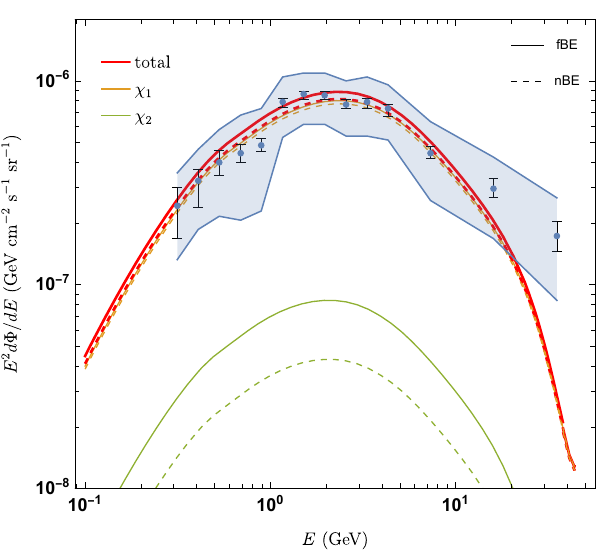}
    \end{subfigure}
    \caption{\textit{Left panel:} Results for an illustrative  scan over masses $M_{\chi_1}$ and $M_s$, keeping fixed $M_{\chi_2}=38$ GeV. The solid and dashed black lines are contours of constant relic abundance reproducing the observed value, from the standard (nBE) and phase-space level (fBE) calculations, respectively. The teal blue shaded contours indicate the region of parameter space most preferred for a DM explanation to the Galactic Centre excess, with the progressively darker shade corresponding to a $60\%$, $90\%$ and $95\%$ C.L. preferred region. The gray lines are contours of the ratio  of nBE abundances $(\Omega h^2)_{\chi_1}/(\Omega h^2)_{\chi_2}$.  And the colour contour plot shows the deviation in the total relic abundance as obtained from the phase-space level solution (fBE) to that from the standard (nBE) calculation, in percentages. \textit{Right panel:}. 
    The differential photon flux for the excess from Galactic Centre,  over a $40^\circ \times 40^\circ$ window, for the parameter space point marked as $\bigstar$ in the left panel. The yellow, green, and red lines show the differential fluxes from $\chi_1$, $\chi_2$, and their sum, respectively, with the  solid (dashed) line giving differential fluxes generated by using the nBE and fBE values of current day abundances. 
    In blue points we show the excess in differential flux from observations\,\cite{Cholis:2021rpp}, along with the associated errors depicted as error bars (statistical) and blue envelope (systematic).}
\label{fig:2}
\end{figure}

In the case of a DM based explanation to the GCE, its spectrum with a peak at few GeV is indicative of the mass of DM sourcing the excess. We therefore direct our attention to the mass range of 20-50 GeV, as has been shown to be preferred in the Coy DM model\,\cite{Boehm:2014hva}. 
Without any loss of generality we can choose $\chi_1$ to be heavier than $\chi_2$ and fix $\chi_2$ mass to fall in this indicated mass range. Then we vary masses $M_{\chi_1}$ and $M_s$ to scan over all the different kinematic possibilities.
We show the results of an illustrative scan in the left panel of fig.\,\ref{fig:2}, where we fix $M_{\chi_2}=38$ GeV, and couplings to $\lambda_{\chi_1}=0.063, \lambda_{\chi_2}=0.0026, \lambda_y=3.86$.  In the figure, the black lines are contours of constant total relic abundance equal to the observed value $\Omega h^2=0.12\pm 0.0012$\,\cite{Planck:2018vyg}, with the dashed and solid lines corresponding to nBE and fBE solutions, respectively. 
The gray lines are contours of the ratio of the (nBE) relic abundances of the two particles, $(\Omega h^2)_{\chi_1}/(\Omega h^2)_{\chi_2}$, showing which particle is the dominant constituent. 

As we change masses  $M_{\chi_1}$ and $M_s$, the parameter space shown comprises of different parts with and without resonant annihilations of $\chi_1$ or $\chi_2$.  From the gray contours of the ratios of relic abundance of the two DS particles, we see that $(\Omega h^2)_{\chi_2}>(\Omega h^2)_{\chi_1}$ for large part of the parameter space. In these regions, the total abundance can then be understood to a large extent as a freeze-out from resonant annihilation ($M_s\simeq 2M_{\chi_2}$) of the dominant $\chi_2$. For each value of $M_{\chi_1} $, we find two values of $M_s$ for which the observed relic abundance is reproduced. This is as expected from a resonant annihilation case, see figure 1 of ref.\cite{Binder:2017rgn}. 
We note that the choice of couplings for the results in this particular figure is such that the conversions are never the leading number changing process for either DS particle. 
So here the conversions have a subleading effect of $\mathcal{O}(10\%)$  on the total abundance (with a possibly larger effect on the subdominant constituent).

To investigate how well the GCE is explained by this double-coy DM model, we look at the teal blue shaded regions in the left panel of fig.\,\ref{fig:2}, with the progressively darker regions showing $60\%, 90\%$ and $95\%$ C.L. preferred regions of the fit, using nBE values for relic densities of $\chi_1$ and $\chi_2$. 
The value of $\chi^2$ is a steep function in the relic density of the DM particle/s sourcing the gamma flux near the best fit regions so the 
visible irregularities on the plot are just an effect of the density of points sampled.

We use \texttt{PPPC4DMID}\cite{Cirelli:2010xx} 
to generate the photon spectrum from the primary final states of heavy quarks, light quarks and leptons, and calculate the differential $\gamma$ ray flux using an NFW density profile with local DM density set to $0.3\, \text{GeV}/\text{cm}^3$ at a distance of $r_0=8.33$ kpc from the Galactic Centre, and a thermally averaged cross section $\langle\sigma v\rangle$ with the DM velocity distribution assumed to follow a Maxwell distribution.  We integrate over the differential flux from the Galactic Centre over a region $40^{\circ} \times 40^{\circ}$. We fit the spectrum of the  excess to the observed value as obtained in the results  of ref.\cite{Cholis:2021rpp}.
We sum over the $\gamma$-ray spectra from $\chi_1$ and $\chi_2$ and as a goodness-of-fit we calculate the reduced $\chi^2$, defined as  
$\chi^2_\text{reduced}\equiv \sum\limits_{i=1}^{N_\text{data}} (th._i^2-obs._i^2) /{(N_\text{data}-1)}$,
where $obs.$ is the observed differential photon flux for the excess from ref.\cite{Cholis:2021rpp}, $th.$ is the  corresponding value calculated for a given parameter space point from the double Coy DM model, and $N_{data}$ is the number of data points.

The flux from annihilation of each DM particle is proportional to the square of its number density, with the total flux equal to their sum. 
For most parts of the parameter space, the contour for the best fit broadly follows a shape parallel to the relic density contour of either $\chi_1$ or $\chi_2$, whichever produces the dominant contribution to the observed flux.
In the left panel of fig.\,\ref{fig:2}, the horizontal part of teal blue shaded best fit region (at $M_s\simeq 84$ GeV) lies parallel to the relic contour since $\chi_2$ forms the dominant relic as well as produces majority of the observed flux. 
The shape of the GCE contour changes along decreasing $M_s$, eventually following the shape of the contour of $\chi_1$ abundance as it comes to dominantly produce the gamma-flux.
The sharp cut at $M_s=2M_{\chi_2}$ is due to  the resonance: as $M_s$ falls  just below the resonant value, the  number density of ${\chi_2}$  increases  drastically while producing too large a flux, and $\chi^2$ increases faster still, which shows up as a sharp cut in its contour.

The overlap of the teal blue shaded regions with the black relic contour line shows the most preferred parameter space for simultaneously producing the observed DM abundance and having a DM sourced GCE explanation. We show in the right panel of fig.\,\ref{fig:2} the spectrum for a point in this overlap, marked as $\bigstar$. In blue points are shown the excess in differential flux from observations, along with their associated errors. The coloured lines show the photon spectrum, for $\chi_1$ in yellow, $\chi_2$ in green and for their sum in red. The dashed and solid lines give the spectrum for each of these with the current day abundance calculated using the standard approach (nBE) and 
from a solution to the phase space level Boltzmann equation (fBE), respectively. We see that the major contribution to the photon flux comes from $\chi_1$, but a non-negligible contribution also comes from the subdominant $\chi_2$. 

Finally, the coloured contours in the left panel of fig.\,\ref{fig:2} show the deviation of fBE solution from nBE solution, specifically $\big((\Omega h^2)_\textrm{fBE})-(\Omega h^2)_\textrm{nBE}\big)/(\Omega h^2)_\textrm{nBE}\times 100$, as indicated by the colour bar.\footnote{The colour contour plots are based on $\sim100$ points each. The deviation in abundances from phase space level effects are often quite sharp in the model parameters, manifesting as kinks in the contours for a sparse grid, as seen here. Thus resolving such spurious features in a two-dimensional scan would require much denser grids, going beyond the CPU limitations. } 
To understand the patterns in this deviation, let's first look at a section with varying $M_s$ at a fixed value of $M_{\chi_1}\simeq 39$\,GeV, where $\chi_2$ is the dominant constituent for all values of $M_s$ shown. For $\chi_2$, the exact resonance lies at $M_s=2M_{\chi_2}=76$ GeV, and we can define a quantity $\delta_2\equiv(2M_{\chi_2}/M_s)^2-1$ to measure how far away from the exact resonance the annihilating particle mass lies. 
We can observe that the deviation of fBE from nBE is as would be expected from previous studies on the deviation of phase space level solutions from the standard calculation for resonant annihilation, see figure 3 in refs.\cite{Binder:2017rgn,Binder:2021bmg}. It is worth noting that  the deviation of fBE from nBE is significant over a broader range of $\delta_2$ (from $-0.2$ to $0.055$), as compared to the previously studied cases with scalar and vector mediated elastic scatterings. This is due to the additional suppression in elastic scattering from the pseudoscalar mediated $t$-channel diagrams, as we highlighted previously, leading to earlier kinetic decoupling and more pronounced effects, underlining the necessity of carrying out a phase space level analysis. 
This relative broadening of deviation is similar to that shown in fig.\,2 of \cite{Binder:2021bmg} with increasing  mass of the scattering partner.

We also observe that the least amount of deviation of fBE from nBE on the contour of the total relic density happens when $\chi_1$ dominates. This is expected at these masses of $\chi_1$, far away from the resonance, with no strong velocity dependence, and no additional suppression of elastic scatterings causing fBE deviation from nBE. It is worth repeating here that we focus on phase space level effects larger than $10\%$, since the CPU limited finite tabulation density and the accuracy settings allow for numerical accuracies only up to a few $\%$. \\

The left panel of fig.\,\ref{fig:2}   with the variation in masses shows a range of kinematics, for a fixed set of couplings. 
However, for a different set of couplings, we would see different or additional features. For example, $\chi_1$ resonant annihilation could dictate the shape of the relic contour if the couplings were such that
$\chi_1$  was the dominant constituent; or for small $\lambda_y$, large $\lambda_{\chi_1},\lambda_{\chi_2}$ one would be in a regime with conversion rates larger than the annihilation rates, and thereby a different set of processes would drive the evolution.
We can equivalently chart these features by zooming into a `perpendicular' direction  to that shown in left panel of fig.\,\ref{fig:2}, by fixing the masses  to a given point in the figure, and scanning over the couplings.
We streamline our discussion, to understand the role of conversions by studying the evolution at the phase space level, by using a parametrization best suited to this purpose.
This can be achieved by parametrizing the couplings in terms of strength of annihilations (to SM particles) and strength of conversions. For any given point one can do a transformation:
\begin{equation}
    \lambda_{\chi_1} \rightarrow c \lambda_{\chi_1} \qquad     \lambda_{\chi_2} \rightarrow c \lambda_{\chi_2} \qquad 
    \lambda_{y} \rightarrow \lambda_{y}/c.
\label{eq:c}
\end{equation}
This changes conversion (grows with $c$), but keeps all else fixed (except the width of the resonance). 
Analogously, one can keep conversions constant while changing annihilations:
\begin{equation}
    \lambda_{\chi_1} \rightarrow a \lambda_{\chi_1} \qquad     \lambda_{\chi_2} \rightarrow  \lambda_{\chi_2}/a \qquad 
    \lambda_{y} \rightarrow \lambda_{y}
\label{eq:a}
\end{equation}
Thus the 3-parameter $\lambda_{\chi_1},\lambda_{\chi_2}, \lambda_{y}$ space can be spanned by these more illustrative parameters $a$ and $c$, with a third parameter giving the overall scale of the couplings. 

In figure \ref{fig:scan_ca} we show the results of two such scans in `conversion-strength' $c$ and `annihilation parameter' $a$, for fixed masses. 
We choose two points from the left panel of fig.\,\ref{fig:2}, which enable us to study various kinematics and interplay of the processes (as discussed in introduction) in the dark matter evolution. Guided by regions of parameter selected by the fit to GCE, we pick the two points marked as $\bigstar$ and $\blacklozenge$  in the left panel of  fig.\,\ref{fig:2}. The point marked as $\bigstar$ has masses $M_{\chi_1}=44\, \textrm{GeV}, M_{\chi_2}=38\, \textrm{GeV}, M_s= 80\, \textrm{GeV}$ and couplings $\lambda_{\chi_1}=\lambda_{\chi_2}=0.05, \lambda_y=1$, with a resonant annihilation of $\chi_2$. While the point marked as $\blacklozenge$ has masses $M_{\chi_1}=39\, \textrm{GeV}, M_{\chi_2}=38\, \textrm{GeV}, M_s= 84\, \textrm{GeV}$ and couplings $\lambda_{\chi_1}=0.063, \lambda_{\chi_2}=0.0026, \lambda_y=3.86$, with resonant annihilation of $\chi_1$ as well as $\chi_2$, and DS particles with nearly degenerate masses. 
We show the results for both in fig.\,\ref{fig:scan_ca}. As in the previous figure \ref{fig:2} (left panel), the black contours demarcate the couplings reproducing the observed relic abundance of $\Omega h^2=0.12 \pm 0.0012$\,\cite{Planck:2018vyg}. The dashed and solid black lines give the relic abundance contours from the standard calculation (nBE) and from the phase-space level solution (fBE), respectively. The teal blue shaded contours show the parameter spaces giving the best fit to the GCE, using the nBE solution for calculating current day relic densities.  Shown by the gray contours are the ratios of relic densities of $\chi_1$ to that of $\chi_2$, as found from the standard calculation (nBE) $(\Omega h^2)_{\chi_1}/(\Omega h^2)_{\chi_2}$. And the gray shaded regions show the parameter spaces where couplings $\lambda_{\chi_1}$ and $\lambda_{\chi_2}$ are larger than 1.\footnote{We do not highlight here the parameter space where $\lambda_y>1$ since the effective coupling dependent on $\lambda_y$ is also multiplied with the SM fermion Yukawa coupling.} The coloured contours show the deviation in total relic abundance as calculated from the phase space level analysis to that from the standard calculation.\\
 
The number changing processes for a given DS particle in the model are one of the following: annihilations to SM $(\chi_\texttt{i},\chi_\texttt{i}\leftrightarrow \textrm{SM},\textrm{SM})$, annihilations to pseudoscalar mediator $(\chi_\texttt{i},\chi_\texttt{i}\leftrightarrow s,s)$ and conversions $(\chi_\texttt{i},\chi_\texttt{i}\leftrightarrow \chi_\texttt{j},\chi_\texttt{j})$ for $\texttt{\small{i}}\neq \texttt{\small{j}}\in\{1,2\}$.  
If the conversion between the DS particles is the leading number changing process for the particle with dominant abundance, the relic abundance of the DS is likely to be ``conversion-driven". In the parameter space of $c,a$ defined above, one tell-tale sign of such a conversion-driven regime will be of the relic density becoming independent of $a$, leading to relic density contours perpendicular to $c$-axis. 
We highlight the features of this conversion-driven freeze-out in subsections\,\ref{sec:phsp1} and \ref{sec:BMdiscuss}, summarising the remaining results in subsection\,\ref{sec:phsp2}. 
In addition to the classification by the type of number-changing process (pattern of chemical decoupling)
we may also look upon the kinetic decoupling  of each DS particle occurring at different times. We organize our discussion within each of the following subsections in accordance to the  specifics of evolution leading to early decoupling and phase space level effects thereby.

\subsection{Discussion of phase space effects in the presence of conversions}
\label{sec:phsp1}
\begin{figure}[t!]
     \centering         \includegraphics[width=\textwidth]{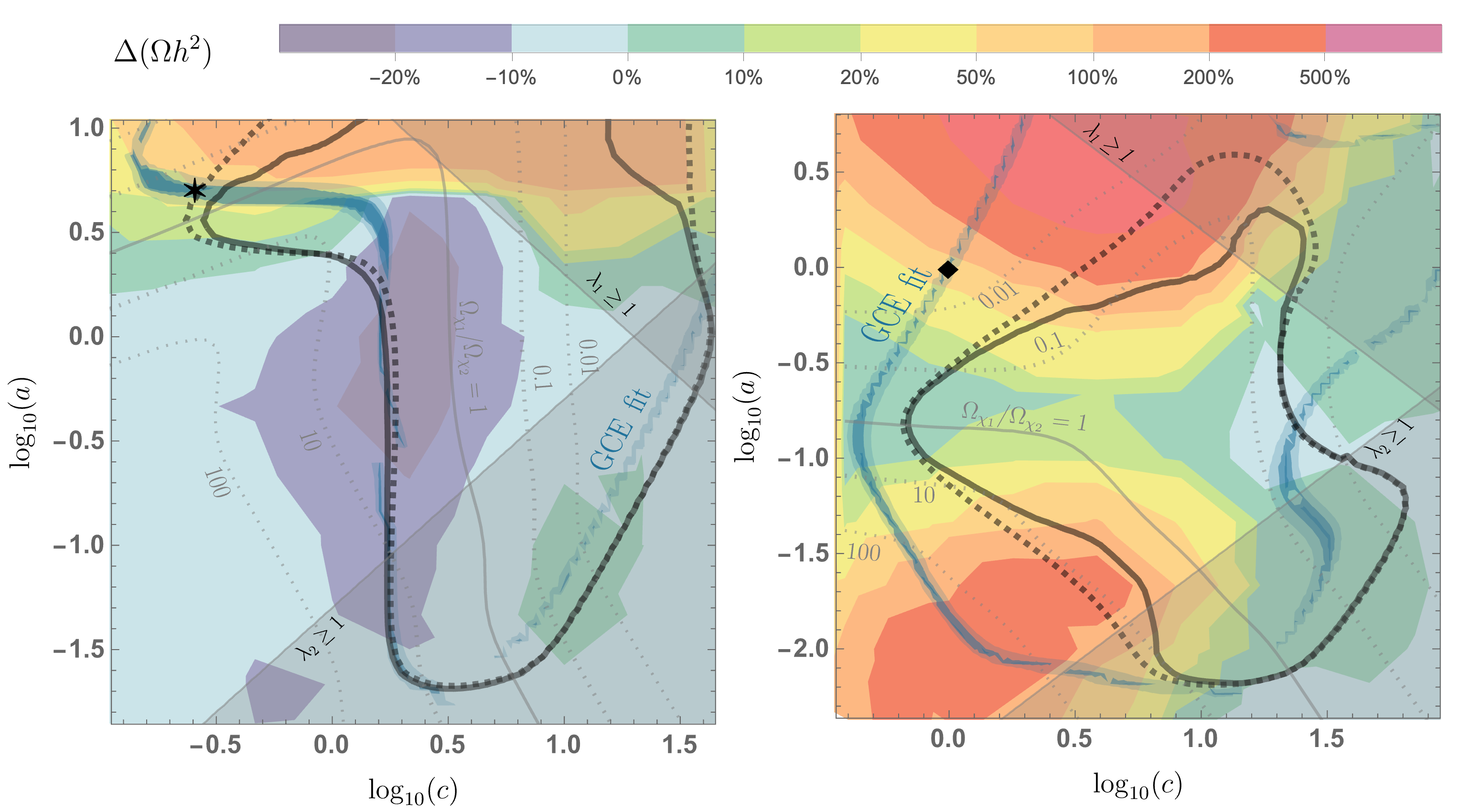}
     \caption{Results for scans over the conversion strength $c$ and annihilation parameter $a$ as defined in eqs.\,\eqref{eq:c} and \eqref{eq:a}, for the point marked in  fig.\,\ref{fig:2} as $\blacklozenge$ and $\bigstar$ shown in the \textit{left} and  \textit{right} panels, respectively. 
     The solid (dashed) black lines are contours of constant relic abundance reproducing the observed value, from the standard (nBE) and phase-space level (fBE) calculations, respectively. The teal blue shaded contours indicate the region of parameter space most preferred for a DM explanation to the Galactic Centre excess, with the progressively darker shade corresponding to a $60\%$, $90\%$ and $95\%$ C.L. preferred region. The gray lines are contours of the ratio  of nBE abundances $(\Omega h^2)_{\chi_1}/(\Omega h^2)_{\chi_2}$.  And the colour contour plot shows the deviation in the total relic abundance as obtained from the phase-space level solution (fBE) to that from the standard (nBE) calculation, in percentages (note that colour coding is different than in fig.\ref{fig:2}). We also show in gray shaded regions the parameter spaces where pseudoscalar mediator couplings to the DS particles start to get large, $\lambda_{\chi_1}, \lambda_{\chi_2}\geq 1$.}
\label{fig:scan_ca}        
\end{figure}

As has been shown in literature, the intertwining of the chemical and kinetic decoupling is captured by the phase space level (fBE) solution, but not by the standard (nBE) calculation which assumes that kinetic decoupling occurs late enough to have no impact on the chemical decoupling. 
This leads to the reported deviation between abundances found from fBE and nBE methods, with the effect being generally significant  when the DM particle cannot be guaranteed to be in kinetic equilibrium over the time period relevant for its freeze-out and there also exists a non-negligible velocity dependence in the cross sections of its leading number changing processes. The former causes non thermal shapes of the phase space distribution to be not washed away instantaneously, while the latter leads to these deviations in DM phase space distribution from the SM equilibrium shape to be carried over to the final relic abundance. 
For example 
in the cases of resonant annihilations\,\cite{Binder:2017rgn,Binder:2021bmg,Duch:2017nbe} and Sommerfeld-enhanced annihilations\,\cite{Binder:2021bmg,vandenAarssen:2012ag,Feng:2010zp}, the annihilation cross sections are enhanced but the elastic scattering cross sections are not enhanced, possibly giving rise to early KD while at the same time the annihilations have a velocity dependence by nature of the enhancements; sub-threshold annihilations (also called `forbidden' DM\,\cite{DAgnolo:2015ujb}) where the elastic scattering partner has a mass comparable to the DM mass with a suppressed number density leading to suppressed elastic scattering rates and thereby early kinetic decoupling, while the annihilations happen only at the high momentum tail of the phase space distribution making it necessarily velocity dependent
\,\cite{Binder:2021bmg,Hryczuk:2022gay,Brummer:2019inq,DAgnolo:2017dbv}. 
In the double Coy DM model, the fBE deviation in the total relic abundance in fig.\,\ref{fig:scan_ca} can  be broadly understood as arising from two of these known scenarios: 
resonant annihilations and sub-threshold annihilations. We organize our discussions on conversion driven freeze-out along these two processes. 

\subsubsection{Conversion driven freeze-out: interplay with resonant annihilation 
}

In the left panel of  fig.\,\ref{fig:scan_ca}, a conversion-driven region can be identified  for values $-1.5\lesssim \textrm{log}_{10}(a) \lesssim 0.3$ at $c\simeq 1.7$, where the relic contour is seen to be almost independent of $a$. 
From the contours of ratio $\Omega_1/\Omega_2$, we see that $\chi_1$ is the dominant relic here, while $\chi_2$ constitutes between $10\%$ and $50\%$ of the total abundance. 
The conversion rate for the dominant particle $\chi_1$ is found to be the leading number changing process which, as discussed above, gives a \textit{conversion driven} freeze-out. 
From the colour contour plot, this same region is also seen to have a negative deviation in the relic abundance from the phase space level computation as compared to the standard computation, which is not explained by a resonant annihilation alone\,\cite{Binder:2021bmg}, and requires more scrutiny that we go into later in sec.\,\ref{sec:BMdiscuss}.
The teal blue shaded contours for parameters giving the best fit for the GCE are, as before, seen to broadly follow the (nBE) relic contour, with the exact shape determined by interplay of current day annihilation cross sections and the number densities for the best fit.\\

In the right panel of fig.\,\ref{fig:scan_ca}, the relic contour can be seen to fall approximately along constant $c$  at $\textrm{log}_{10}(c)\simeq 1.3, \textrm{log}_{10}(a)\simeq -0.48$, where $\chi_2$ forms the majority of the total relic abundance. From the colour contours, we also see that the standard computation underestimates the relic abundance, which can be explained by the velocity-dependence in conversion process from resonance on $s$. In this parameter space, the rates of conversions are larger than the annihilation rates for most of the relevant times, for both $\chi_1$ and $\chi_2$. \\

\subsubsection{Conversion driven freeze-out: interplay with sub-threshold annihilation }
Dark matter freeze-out can be governed by annihilation processes that are inaccessible in the limit of vanishing velocity, but are made accessible at finite temperatures. These sub-threshold processes lead to a necessarily velocity dependent number changing process.
Of the possible number changing processes in our results, $\chi_\texttt{i},\chi_\texttt{i}\leftrightarrow \textrm{SM,SM}$, $\chi_\texttt{i},\chi_\texttt{i}\leftrightarrow s,s$ and $\chi_\texttt{i},\chi_\texttt{i}\leftrightarrow \chi_\texttt{j},\chi_\texttt{j}$ where $(\texttt{\small{i}}\neq \texttt{\small{j}}\in\{1,2\})$, only the conversions can lead to a sub-threshold annihilation. The choice of pseudoscalar masses in both panels of  fig.\,\ref{fig:scan_ca} inhibit  efficient sub-threshold annihilations $M_s>M_{\chi_1},M_{\chi_2}$.
While significant coupling to SM  fermions usually allows for some efficient annihilation channel uninhibited by a threshold. 
Thus the case of a conversion driven freeze-out with  conversions  $\chi_2,\chi_2\leftrightarrow \chi_1,\chi_1$ being the leading number changing process and $\chi_2$ being the dominant component, along with the two DS particles having similar masses ($M_{\chi_1}\gtrsim M_{\chi_2}$), leads to a (necessarily conversion-driven) freeze-out with sub-threshold annihilations. 

The choice of masses in the right panel of fig.\,\ref{fig:scan_ca} makes this a possibility in the  region of parameter space that we  previously highlighted to have conversion-driven freeze-out. The leading number changing process -conversions- could in addition to the resonance on $s$ , also have a velocity dependence from sub-threshold annihilations $\chi_2,\chi_2\rightarrow\chi_1,\chi_1$.
It is, however, not possible to disentangle the two effects, since both the resonant and sub-threshold annihilations would present as a similar effect in the difference from fBE compared to nBE. 

To study the process of freeze-out from sub-threshold conversions, we direct our attention to a parameter space where the sub-threshold conversion is the leading number changing process as well as its \textit{only} source of velocity-dependence. 
\begin{figure}[t!]
\centering
\includegraphics[scale=0.6]{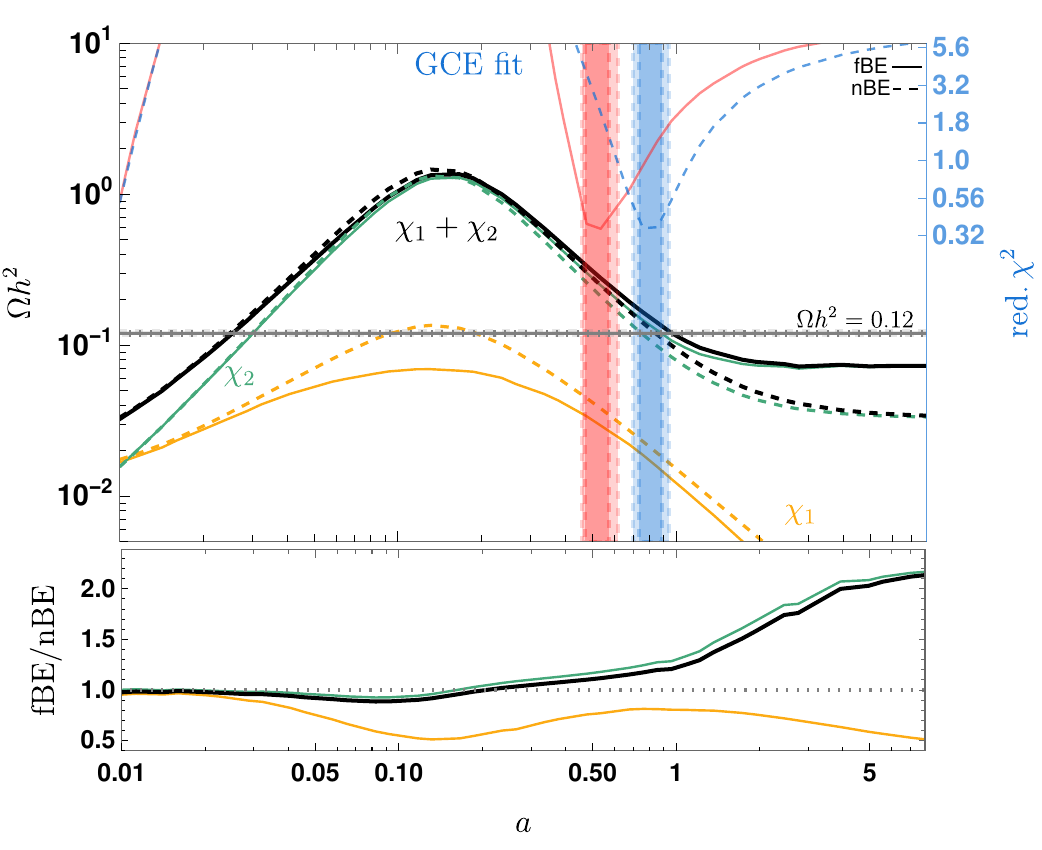}
\caption{Results for a scan over annihilation parameter $a$ as defined in eq.\,\eqref{eq:a}, for sub-threshold conversions, $M_{\chi_1}=31$ GeV, $M_{\chi_2}=30.6$ GeV, $M_s=75$ GeV. In the \textit{top panel}, the yellow, green and black lines show the relic abundances for $\chi_1$, $\chi_2$ and their sum, respectively, with the solid (dashed) line showing the values from fBE (nBE) solutions. The gray line marks the value of the observed relic abundance $\Omega h^2=0.12 \pm 0.0036$. The reduced $\chi^2$ from fits to the GCE are shown by the solid red and blue dashed lines using current day abundances as obtained from fBE and nBE calculations, respectively. We show the $a$-values preferred at $60\%$ and $90\%$ C.L in the shaded coloured regions, to highlight the difference in preferred parameters as obtained from the standard calculation and the improved phase space level analysis. In the \textit{bottom panel} are given the ratios of abundances from fBE to nBE, to highlight the size of the effects.}
\label{fig:subth}
\end{figure}

In fig.\,\ref{fig:subth} we show the results for an example case with masses: $M_{\chi_1}=31$ GeV, $M_{\chi_2}=30.6$ GeV, $M_s=$ 75 GeV and  couplings: 
$\lambda_{\chi_1}^0=1.82, \lambda_{\chi_2}^0=0.019, \lambda_{y}=0.22$. For fixed conversion strength $c$, we scan over the annihilation-parameter $a$ (similar to the vertical section in the right panel of  fig.\,\ref{fig:scan_ca} where we noticed the sub-threshold conversion driven freeze-out, only with masses far away from the resonance). Note that the coupling $\lambda_{\chi_1}$ increases and $\lambda_{\chi_2}$ decreases with increasing $a$, so the annihilation and scattering cross-sections for $\chi_1\,\,(\chi_2)$ decrease (increase) as the value of annihilation-parameter $a$ increases, while the conversion cross-section remains constant. 
In fig.\,\ref{fig:subth}, the coloured lines show the relic densities of $\chi_1$ in yellow, $\chi_2$ in green and their sum in black, with the solid (dashed) line for each colour giving the fBE (nBE) solutions. 
Shown in blue are the reduced $\chi^2$ values from the fit to the GCE, with the blue dashed and solid red lines giving the values obtained by using the relic abundance values from nBE and fBE solutions, respectively.  The blue and red shaded regions show the values of $a$ that are preferred by the GCE fit at $90\%$ and $95\%$ C.L., from the nBE and fBE solutions. The gray line marks the observed abundance with the width giving $3\sigma$ errors.

At large values of $a$, we observe the tell-tale conversion-driven feature, independence on $a$.  
And we see that for $\chi_2$  which freezes out via the sub-threshold conversions, density from fBE is larger than that from nBE, as expected from literature\,\cite{Binder:2021bmg}.
Moving our attention to the rest of the curve, we may  understand the dependence on $a$ of the abundances found from nBE as follows. With increasing $a$, the relative strengths of the rates for conversions and annihilations of the two DS particles over the relevant times changes broadly as, 
\begin{equation}
\label{eq:rates_subth}
    \begin{array}{cll}
    \begin{array}{c}
    \Gamma_{2,\text{ann.}} > \Gamma_\text{convs.} >\Gamma_{1,\text{ann.}}     \\ 
    \\
    \Gamma_\text{convs.} > \Gamma_{2,\text{ann.}} \sim \Gamma_{1,\text{ann.}} \\
    \\
    \Gamma_{1,\text{ann.}} >\Gamma_\text{convs.} >\Gamma_{2,\text{ann.}} 
    \end{array} 
         &
    \begin{tikzpicture}[overlay]
    \node (A) at (0, -1) {};
    \node (B) at (0, 1.3) {};
    \draw[->, to path={-| (\tikztotarget)}]
    (A) edge (B);
    \end{tikzpicture}        
        &
    \rotatebox[origin=c]{-90}{\text{\textit{increasing} $a$}}    
    \end{array}
\end{equation} 

where the two conversion rates are nearly equal because of the nearly degenerate masses of $\chi_1$ and $\chi_2$ $(M_{\chi_1}\gtrsim M_{\chi_2})$, and we club them together in $\Gamma_{\text{convs.}}$. This rate of conversions is nearly unchanged as $c$ remains fixed.
Taken together with eq.\,\eqref{eq:rates_subth}, we see that the sum of number changing processes first increases and then decreases, with increasing $a$, and conversions are found to be the leading process around $a=0.1$. The dominant constituent changes from $\chi_1$ to $\chi_2$, also as expected from the rates from eq.\,\eqref{eq:rates_subth}.

For the phase space level analysis, we  show the ratio of the abundances for each of the particles and their sum, as obtained from fBE to that obtained from nBE, in the lower panel.
Interestingly, we find that the fBE solution gives a \textit{smaller} value for the relic abundance of $\chi_1$, with this effect being maximized when conversion are the leading number changing process, which being sub-threshold leads to an early kinetic decoupling and a faster cooling of $\chi_2$ distribution in fBE than in nBE. The conversions  $(\chi_2,\chi_2\rightarrow\chi_1,\chi_1)$  from the cooler $\chi_2$ distribution  would therefore be further suppressed in fBE as compared to  in nBE, leading towards the \textit{decrease} in $\chi_1$. The dominant contribution to the fluxes at Galactic Centre come from $\chi_2$ for the majority of the parameter space shown in fig.\,\ref{fig:subth}. 
Thus the significant deviation in $\chi_1$ relic abundance from fBE to nBE, even though does not lead to a large deviation in the total abundance, changes the preferred parameter space for the explanation to the GCE with the preferred $a$ values from nBE and fBE shown by the blue and red shaded bars, respectively.
This is an example of how even when accounting for  phase space level effects might seemingly only lead to a small correction in the total relic abundance, 
can significantly change the preferred region for GCE fit.

\subsection{Discussion of remaining phase space level effects }
\label{sec:phsp2}
To complete the analysis of phase space level effects in a two-component DM model with conversions, we discuss in this section the remaining effects as seen in fig.\,\ref{fig:scan_ca}, which haven't been covered in the previous sections, where the impact of conversions is less severe and/or it is not possible to disentangle the various processes enough to clearly argue for it to be so.  We discuss in the following the deviations of fBE from nBE for each figure.

In fig.\,\ref{fig:scan_ca} (left panel), we find increasingly large deviation of fBE from nBE results, with increasing $a$ for $\log_{10}(a)\gtrsim 0.5$. The nBE contour can be seen to follow two distinct shapes for small and large values of $c$. For small $c$, $\lambda_y$ is large and $s$ decays dominantly to SM with the $\chi_2$ resonant annihilation cross section scaling as 
$$\sigma_{\chi_2,\chi_2\rightarrow\textrm{SM},\textrm{SM}} \propto \lambda_{\chi_2}^2\Gamma_{s\rightarrow\textrm{SM}}/\Gamma_s^2\propto \lambda_{\chi_2}^2/\lambda_y^2 \propto c^4/a^2.$$ 
While for large $c$, $\lambda_y$ is small  and $s$ decays dominantly to $\chi_2$ so the $\chi_2$ resonant annihilation cross section scales as
$$\sigma_{\chi_2,\chi_2\rightarrow\textrm{SM},\textrm{SM}} \propto \lambda_{\chi_2}^2 \Gamma_{s\rightarrow\textrm{SM}}/\Gamma_s^2\propto \lambda_y^2\propto 1/c^2,$$
becoming independent of annihilation parameter $a$. However, this not a case of conversion-driven freeze-out.  Since the elastic scattering rates continue to scale $\propto \lambda_{\chi_2}^2\propto 1/a^2$, kinetic decoupling occurs progressively earlier as $a$ increases and the fBE effect increases. 
Both these effects would only be seen for resonant annihilations.

We briefly note the parts of the parameter space where the fBE and nBE results match well in fig.\,\ref{fig:scan_ca}: where $\chi_1$ form the dominant relic, $\log_{10}(a)\simeq 0, \log_{10}(c)\simeq -0.5$, since  the leading number changing process of its annihilation is non-resonant and does not carry any strong velocity dependence; and where $\chi_2$ forms the dominant relic, $-2\leq \log_{10}(a)\leq 0, 0.5\leq \log_{10}(c)\leq 1.6$, so the relic contour falls along contours of constant $\sigma v_{2,2\rightarrow SM,SM}$ but the elastic scattering rates are large enough for there not to be any significant phase space level effects.  

In the right panel of fig.\,\ref{fig:scan_ca} the effect of conversions and annihilations are more difficult to disentangle since the two DS particles have similar masses and couplings. For small $c$ where conversions are weak, the deviation of fBE results from nBE can be understood as arising from the resonant annihilations of $\chi_1$ or $\chi_2$, whichever is the dominant relic. For larger values of $c$ the rates of conversion and annihilation  become comparable and the fBE deviation from nBE is a result of the convolution of these processes.\\

We now return to the benchmark point from section\,\ref{sec:BM} exhibiting non-thermal evolution in phase-space densities and discuss the results for the DM particle abundances and phenomenology in the form of impact on the gamma ray signal from the Galactic Centre. Together with the observations on conversion-drive freeze-out, we find that the benchmark point is qualitatively similar to the resonant conversion-driven freeze out identified in left panel of fig.\,\ref{fig:scan_ca}.  
We carry out a phase-space analysis along  changing conversion strength $c$ and look closely at the phenomenological consequences thereby. 

\subsection{Benchmark -- phenomenological consequences of departure from LTE}
\label{sec:BMdiscuss}
\begin{figure}[t!]
    \centering
    \begin{subfigure}[b]{0.495\textwidth}
    \includegraphics[width=\textwidth]{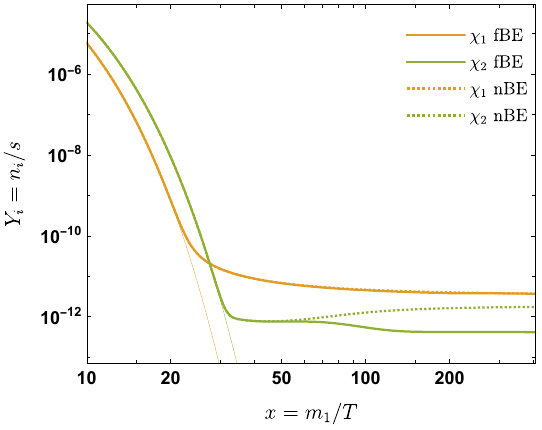}
    \end{subfigure}
    \begin{subfigure}[b]{0.483\textwidth}
    \includegraphics[width=\textwidth]{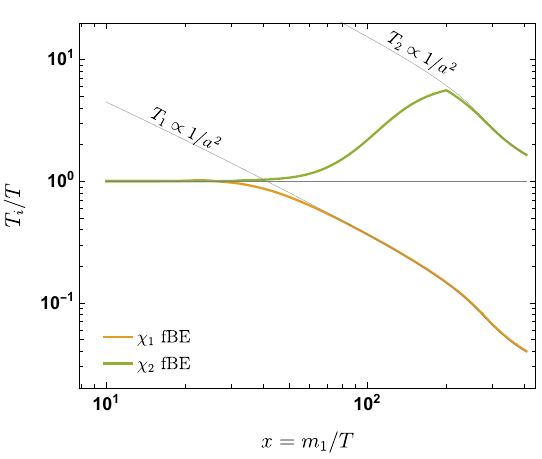}
    \end{subfigure}
    \caption{Evolution of $\chi_1$ and $\chi_2$ for the benchmark point. \textit{Left panel:} Yields of DM particles, $Y_{\chi_\texttt{i}}\equiv n_{\chi_\texttt{i}}/s$ where  $n_{\chi_\texttt{i}}\equiv g_{\chi_\texttt{i}}/(2\pi)^3\int \text{d}^3\vec{p} f_{\chi_\texttt{i}}(p,x)$ is the number density and  $s$ is the entropy density. \textit{Right panel:} Effective temperature of DM particles $T_{\texttt{i}}\equiv \frac{g_{\chi_\texttt{i}}}{n_\texttt{i}(x)}\int \frac{\text{d}^3\vec{p}}{(2\pi)^3}\frac{|\vec{p}|^2}{3E_\texttt{i}}f_{\chi_\texttt{i}}(p,x)$, scaled with the SM bath temperature $T$. In the thin gray lines we show the shape of an evolution proportional to $1/a^2$, where $a$  is the expansion scale factor, to highlight the point of kinetic decoupling after which the DM particles  evolve freely under expansion. This also serves to highlight the rather significant heating of $\chi_2$ before the eventual kinetic decoupling (see text for description).  
    }
\label{fig:BMYy}
\end{figure}

We started our discussion from showing in section\,\ref{sec:BM} the evolution of the shape of the distribution function for a selected benchmark. Let us finish by circling back to explore this point further. In fig.\,\ref{fig:BMYy} we show the evolution of the yields (left panel) and the effective temperature (right panel) for this particular benchmark. 
The heavier state $\chi_1$ has smaller couplings to the SM, leading to its abundance after chemical decoupling being larger than the one of $\chi_2$. It follows that the rate of $\chi_1 \chi_1 \rightarrow \chi_2 \chi_2$ conversions is large enough to have a significant impact on the distribution and evolution of $\chi_2$ particles, as seen also in section\,\ref{sec:BM}. Surprisingly, though, this additional influx of $\chi_1$ particles leads here to a \textit{decrease} of its final abundance. This is caused by the interplay of conversions and resonance annihilation of $\chi_2$, which so happens to be most efficient for particles in the momentum range that is additionally fuelled by conversions.\footnote{Analogous effect was found in \cite{Hryczuk:2022gay} where the injection of DM particles came from a decay of a heavier DS state, and the increase in annihilation strength came from a kinematical threshold.}
Even though the increased annihilation deplete its number density, $\chi_2$ population ends up significantly heated up, being a factor of a few, compared to the case if it stayed in kinetic equilibrium.

\begin{figure}[t!]
    \centering
    \includegraphics[width=0.7\linewidth]{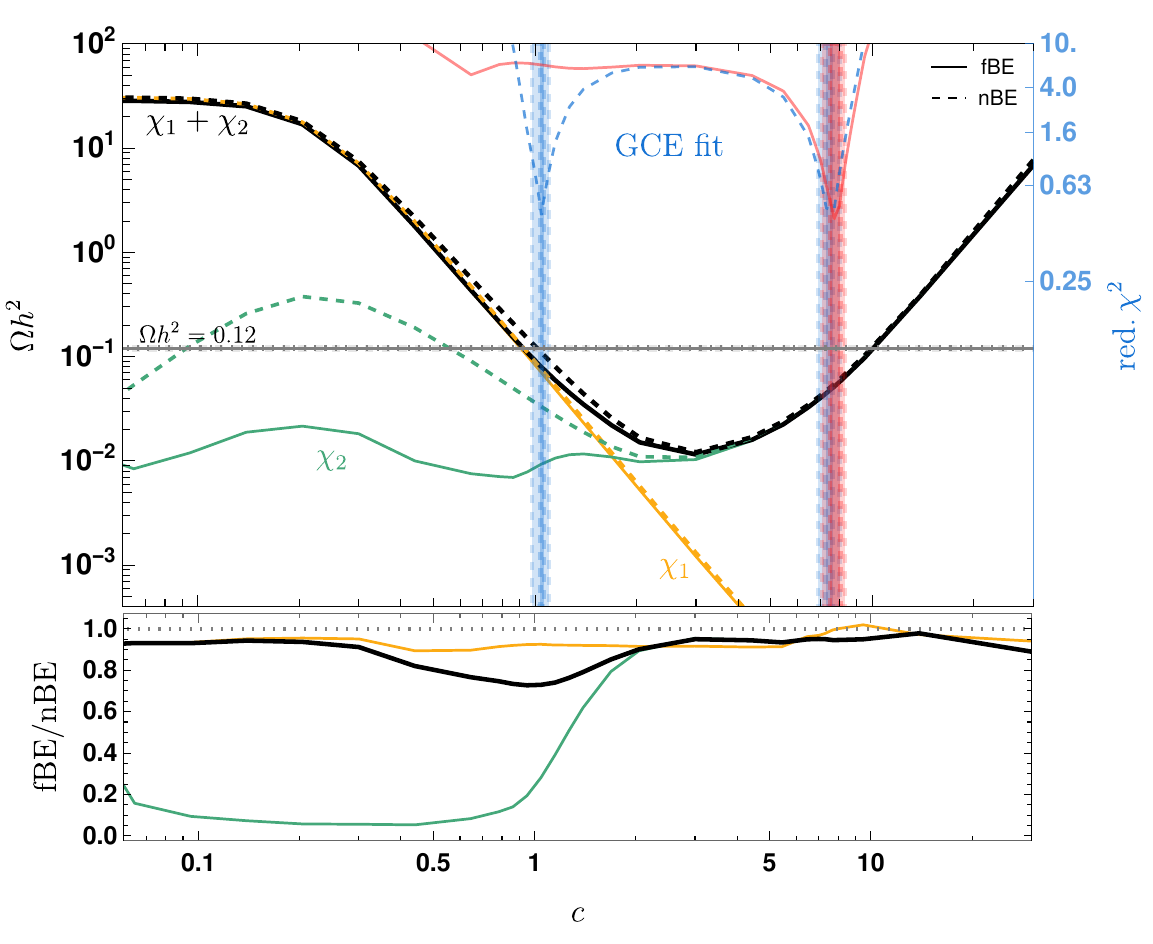}
    \caption{
    Results for a scan over conversion strength $c$ as defined in eq.\,\eqref{eq:c} for the benchmark point $M_{\chi_1} = 44 $ GeV, 
    $M_{\chi_2} = 38 $ GeV, 
    $M_s = 80 $ GeV, showcasing interplay of conversions and resonant annihilations. The coloured lines and shaded regions are same as fig.\,\ref{fig:subth}.
    }
\label{fig:BM1D}
\end{figure}

To see how such an effect depends on the efficiency of conversion process, we scan over the conversion strength $c$ starting with the benchmark point at $c=1$, and show the results in fig.\,\ref{fig:BM1D}. For $c\leq 1$, we find that $\chi_1$ continues to be the dominant DM component. As $c$ increases its total number changing rate comes to be dominated by conversions leading to smaller final abundance, and $\chi_2$ becoming the dominant relic.
This in turn means the conversions from $\chi_1$ to $\chi_2$ do not have any noticeable effect on $\chi_2$ abundance at large $c$. Additionally, $\lambda_{\chi_2}$ increases with increasing $c$, leading to an increasingly large partial decay width of $s$ to $\chi_2$ and thereby a larger total decay width of $s$. 
As discussed in ref.\,\cite{Binder:2021bmg}, the deviation of fBE abundances from the nBE abundances is smaller for a wider  resonance, hence the overlap of fBE and nBE results at large $c$. \\

\begin{figure}[t!]
    \centering   \includegraphics[scale=0.85]{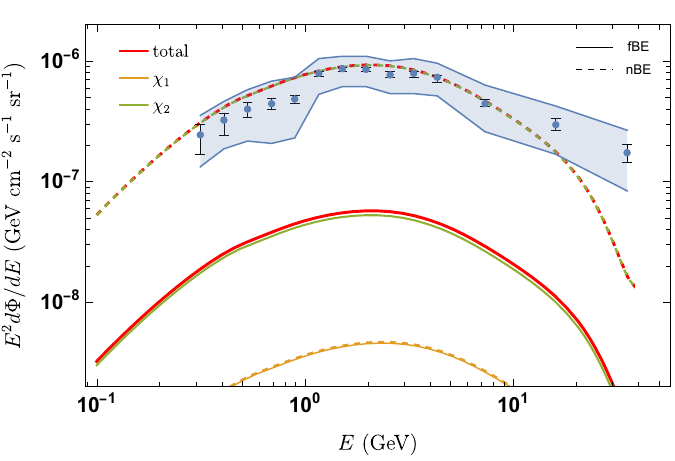}
    \caption{ The differential photon flux for the excess from Galactic Centre,  over a $40^\circ \times 40^\circ$ window, for the benchmark point from sec.\,\ref{sec:BM}. The yellow, green, and red lines show the differential fluxes from $\chi_1$, $\chi_2$, and their sum, respectively, with the solid (dashed) line giving differential fluxes generated by using the fBE (nBE) values of current day abundances. 
    In blue points we show the excess in differential flux from observations\,\cite{Cholis:2021rpp},  along with the associated errors depicted as error bars (statistical) and blue envelope (systematic).}
\label{fig:BM1_GCE}
\end{figure}

For $c\simeq 1$, which is the same as the initial example point, the two particles have comparable abundances and so the fBE deviation from nBE carries over to the total abundance, giving $\sim 20\%$ effect. 
More significantly impacted by this phase space level analysis is the GCE fit though. Using the relic abundances from the standard calculation, we find two parameter space points that give a good 
fit to the observed GCE. We highlight this in fig.\,\ref{fig:BM1D} with the blue shaded bars. The corresponding parameter space one arrives at by using the fBE solution though is shown via red shaded regions. We find that the previously preferred parameter space points at $c\simeq 1$ does not lead to a good fit to the GCE. 
We show the spectrum for the point corresponding to $c=1$ in fig.\,\ref{fig:BM1_GCE}. 
The total abundance is $\Omega h^2=0.126$ (nBE) and $0.097$ (fBE) while the fits give reduced $\chi^2=0.51$ (nBE) and $62$ (fBE). 
The present day cross sections are $\langle\sigma v\rangle_0^{\chi_1}=4.34 \cdot 10^{-28}$ cm$^3/$s and $\langle\sigma v\rangle_0^{\chi_2}=4.46 \cdot 10^{-25}$ cm$^3/$s, and the total flux is dominantly produced by $\chi_2$, so that the larger deviation in the sub-dominant component of the abundance can indeed be significant in its impact on another observable, here the gamma ray flux from Galactic Centre.    

\section{Conclusions}
\label{sec:conclusions}

In a generic multi-component dark sector, kinetic equilibrium between the SM plasma and all the particles whose dynamics is relevant for the freeze-out process is not at all guaranteed. Nevertheless, majority of the studies found in literature, as well as state-of-the-art public numerical tools applicable to multi-component scenarios, do rely on the assumption of kinetic equilibrium. In this work, we have extended DRAKE -- a code designed to provide solutions for the relic abundance including non-equilibrium effects -- to efficiently solve the Boltzmann equation for the phase space distributions of two dark sector states, including all relevant annihilation, scattering and conversion processes.

To exemplify the impact of said departure from kinetic equilibrium, we have performed a detailed analysis of a specific model of phenomenological interest, being a minimal extension of the Coy DM scenario. The model comprises of visible SM sector and a dark sector containing two fermions and a pseudoscalar mediator. As in the original Coy DM, the studied realisation can at the same time account for all the observed DM, fit well the excess in gamma rays as observed from the GC and avoid limits from direct detections as well as collider searches. At the price of including one more state, the analysed model leads to more involved freeze-out dynamics, in particular allowing for a conversion-driven realisation consistent with the observed GCE signal.

The significant role that conversions play in this model gives an opportunity to study in detail their impact on the evolution of the dark sector particles at the phase space level. We show that there is a non-trivial interplay between conversions and annihilations that in a significant way depends on the kinetic equilibration or lack thereof. 

Focusing on the parameter regions that lead to the observed relic abundance and provide a good fit to GCE, we find that departure from kinetic equilibrium can alter the predictions for the total DM abundance by more than 100\%, while  being in the range from around -20\% to 50\% in most of the interesting parameter space. Additionally, when looking at the abundances of the two DM constituents separately, the observed effects can be vastly larger, $\mathcal{O}(\text{few})$, and even up to an order of magnitude. It is worth emphasizing that such a large modification of the abundance of a subdominant component can  affect the expected present-day gamma ray flux in a significant way, and as shown with explicit examples, completely change the preferred region for the GCE fit.

The delineation of the precise impact of conversions and other relevant processes on the departure from kinetic equilibrium, distorting the shape of the momentum distribution, and finally evolution and the resulting relic abundance is rather difficult, as well as is predicting the numerical size of the error one does by assuming full kinetic equilibrium. So is predicting the  error incurred by assuming full kinetic equilibrium. Thus, this work also provides an implementation of the numerical framework of including conversions in a generic two-component DM model, that is going to be publicly available in the next release of DRAKE.

Finally, let us note that the presented analysis was done on a realisation where both dark sector states were stable, which does not have to be the case in other models. The resulting decay processes provide additional source of potential violation of kinetic equilibrium and are left for future work.

\section*{Acknowledgments}

We would like to cordially thank Tobias Binder and Ayuki Kamada for helpful discussions. 
This work was supported by the National Science Centre, Poland, under the research grant No. 2021/42/E/ST2/00009.

\begin{appendix}
\section{On the limitations of Fokker-Planck approximation}
\label{sec:app}
In a typical scattering event of a heavier, non-relativistic DM particle against a lighter, relativistic particle from the SM bath, the momentum transfer is typically smaller than the incoming momentum of the DM particle. The Fokker-Planck (FP) approximation is arrived at by expanding in this small quantity $|\Delta \vec{p}\,|/|\vec{p}_\text{DM}\,|$ and small DM velocity $|\vec{p}_\text{DM}|/E_\text{DM}$, and keeping only the leading order terms in the expansion\cite{Bringmann:2006mu}. This expansion was improved upon by including the leading relativistic corrections\cite{Binder:2021bmg,kasahara2009neutralino,gondolo2012effect,Binder:2016pnr} which we give in eq.\,\ref{eq:FP}. 
The difficulty with the elastic scattering collision term lies in the integrations over the unknown DM distribution functions. 
After the expansion and a collection of the leading order terms as described above, we arrive at a simplified  expression with two parts. First, a momentum transfer rate $\gamma$ which carries the only integration and that over equilibrium distribution functions of particles in the heat bath allowing for it to be pre-tabulated over $x$. And a second part with the FP-like operator with derivatives acting on the unknown DM distribution function. Thus avoiding having to do any repeated, costly integrations at each step of the solver. It is worth repeating that the Fokker-Planck operator is manifestly number conserving and admits the  Maxwellian distribution as a stationary solution, as indeed any elastic scattering collision term  must.

 \begin{figure}[t]
     \centering
     \includegraphics[width=0.75\linewidth]{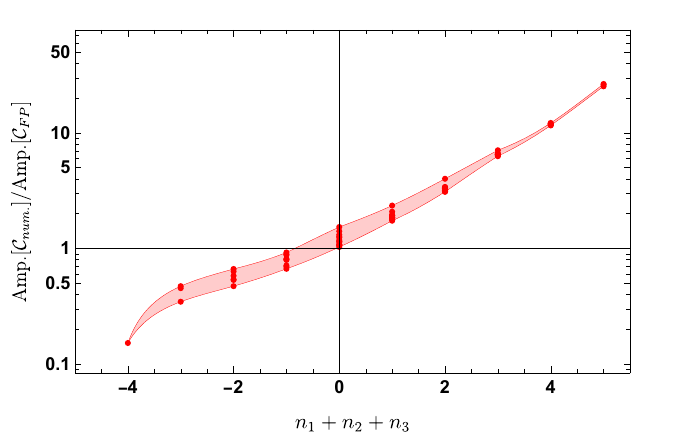}
     \caption{The ratio between the amplitude of the elastic scattering collision term and the one calculated in the Fokker-Planck approximation for amplitude of the form as in eq. \ref{eq:123} with $m_\textrm{SM}/m_\text{DM}=1.1$ and $T_\text{DM}/T_\textrm{SM}=0.95$. 
     For any given value of $n_1+n_2+n_3$ there are a few combinations of $n_1$, $n_2$ and $n_3$ that are given as red dots, while the red envelope is given to guide the eye.}
     \label{fig:App}
 \end{figure}
 
However, as has been pointed out before\cite{Binder:2021bmg,Ala-Mattinen:2022nuj,Hryczuk:2022gay}, the expansion and hence the approximation has limitations and fails for example in the limit of a scattering particle of mass comparable to the DM mass. In fact the approximation fails as soon as any of the two quantities expanded in $\Delta \vec{p}/\vec{p}_\text{DM}$ and $|\vec{p}_\text{DM}|/E_\text{DM}$,  ceases to be \textit{small}. We find by studying the consecutive terms of the expansion for an example case of heavy scattering partner, that indeed the sum of the  consecutive terms of the expansion no longer converge. 
Given the time efficiency of the Fokker-Planck approximation of the elastic scattering collision term, we find it worthwhile to report the conditions under which the approximation fails, necessitating the evaluation of the full collision term with costly but unavoidable numerical integrations. The  FP approximation is found to work well as long as:
\begin{enumerate}
    \item Scattering particle mass is smaller than DM mass.
    \item DM temperature is not very different from the SM temperature.
    \item Elastic scattering amplitude doesn't have a strong dependence on momentum transfer and velocities in the process (so there is less sensitivity to the dropping of higher order terms)
\end{enumerate}
When \textit{all} of the above conditions hold true simultaneously, as in a canonical WIMP scattering, the FP approximation is very good, matching the numerically obtained collision term at the level of a few $\%$. While if \textit{any} one of the conditions do not hold true, the goodness of the FP approximation is no more guaranteed and one is required to evaluate the numerical collision term as discussed in section\,\ref{sec:elsc}.
The last of these conditions is dependent only on the  type of interaction in the particle physics model and can be made more explicit by expanding the squared amplitude for scattering in terms of the transfer momentum $t$, the relative velocity $(s-(m_\text{DM}+m_\textrm{SM})^2)$ and the combination of velocities $(u-(m_\text{DM}-m_\textrm{SM})^2)$ as follows
\begin{equation}
|\mathcal{M}|_{\text{DM},\textrm{SM}\leftrightarrow \text{DM}, \textrm{SM}}^2 \longrightarrow 
t^{n_1}\left(s-(m_\text{DM}+m_\textrm{SM})^2\right)^{n_2} \left(u-(m_\text{DM}-m_\textrm{SM})^2\right)^{n_3}.
\label{eq:123}
\end{equation}
For  $n_1+n_2+n_3=0$, the amplitude is not very sensitive to momentum-transfer and velocity  so the errors accrued from dropping of the higher order terms in the FP expansion are  small. While for all other values, we might expect for the errors to be larger. We demonstrate this in fig.\,\ref{fig:App} where we show the error in using the FP approximation in lieu of the full numerical collision term, in terms of the ratio of the amplitudes of the collision terms from the two methods. We see that the ratio can be quite different from 1 when $n_1+n_2+n_3\neq 0$ given the  mass ratio $m_\textrm{SM}/m_\text{DM}=1.1$ and ratio of effective DM temperature to the SM bath temperature  $T_\text{DM}/T_\textrm{SM}=0.95$. It is worth stressing, though, that typically the elastic scattering rate around freeze-out is a steep function of the temperature, leading to a somewhat suppressed sensitivity to the amplitude of the elastic scattering collision term. Therefore, the Fokker-Planck approximation often leads to approximately correct results even if the aforementioned ratio is not very close to 1. Nevertheless, for higher accuracy the numerical calculation is required.
\end{appendix}

{\small
\bibliography{biblio}

\providecommand{\bysame}{\leavevmode\hbox to3em{\hrulefill}\thinspace}
\frenchspacing
\newcommand{\origttfamily}{}
\let\origttfamily=\ttfamily
\renewcommand{\ttfamily}{\origttfamily \hyphenchar\font=`\-}

\begin{thebibliography}{10}

\bibitem{Arcadi:2024ukq}
G.~Arcadi, D.~Cabo-Almeida, M.~Dutra, P.~Ghosh, M.~Lindner, Y.~Mambrini, J.~P. Neto, M.~Pierre, S.~Profumo, and F.~S. Queiroz, \texttt{arXiv:2403.15860} [hep-ph].

\bibitem{Gondolo:1990dk}
P.~Gondolo and G.~Gelmini, Nucl. Phys. B \textbf{360} (1991), 145.

\bibitem{Edsjo:1997bg}
J.~Edsjo and P.~Gondolo, Phys. Rev. D \textbf{56} (1997), 1879, \texttt{hep-ph/9704361}.

\bibitem{Aboubrahim:2023yag}
A.~Aboubrahim, M.~Klasen, and L.~P. Wiggering, JCAP \textbf{08} (2023), 075, \texttt{arXiv:2306.07753} [hep-ph].

\bibitem{Duch:2017nbe}
M.~Duch and B.~Grzadkowski, JHEP \textbf{09} (2017), 159, \texttt{arXiv:1705.10777} [hep-ph].

\bibitem{Binder:2017rgn}
T.~Binder, T.~Bringmann, M.~Gustafsson, and A.~Hryczuk, Phys. Rev. D \textbf{96} (2017), no.~11, 115010, \texttt{arXiv:1706.07433} [astro-ph.CO], [Erratum: Phys.Rev.D 101, 099901 (2020)].

\bibitem{Binder:2021bmg}
T.~Binder, T.~Bringmann, M.~Gustafsson, and A.~Hryczuk, Eur. Phys. J. C \textbf{81} (2021), 577, \texttt{arXiv:2103.01944} [hep-ph].

\bibitem{Liu:2023kat}
Y.~Liu, X.~Liu, and B.~Zhu, Phys. Rev. D \textbf{107} (2023), no.~11, 115009, \texttt{arXiv:2301.12199} [hep-ph].

\bibitem{Du:2021jcj}
Y.~Du, F.~Huang, H.-L. Li, Y.-Z. Li, and J.-H. Yu, JCAP \textbf{04} (2022), no.~04, 012, \texttt{arXiv:2111.01267} [hep-ph].

\bibitem{Abe:2021jcz}
T.~Abe, Phys. Rev. D \textbf{104} (2021), no.~3, 035025, \texttt{arXiv:2106.01956} [hep-ph].

\bibitem{Fitzpatrick:2020vba}
P.~J. Fitzpatrick, H.~Liu, T.~R. Slatyer, and Y.-D. Tsai, Phys. Rev. D \textbf{106} (2022), no.~8, 083517, \texttt{arXiv:2011.01240} [hep-ph].

\bibitem{Ala-Mattinen:2022nuj}
K.~Ala-Mattinen, M.~Heikinheimo, K.~Kainulainen, and K.~Tuominen, Phys. Rev. D \textbf{105} (2022), no.~12, 123005, \texttt{arXiv:2201.06456} [hep-ph].

\bibitem{Filimonova:2022pkj}
A.~Filimonova, S.~Junius, L.~Lopez~Honorez, and S.~Westhoff, JHEP \textbf{06} (2022), 048, \texttt{arXiv:2201.08409} [hep-ph].

\bibitem{Chang:2022jgo}
C.~Chang, P.~Scott, T.~E. Gonzalo, F.~Kahlhoefer, A.~Kvellestad, and M.~White, Eur. Phys. J. C \textbf{83} (2023), no.~3, 249, \texttt{arXiv:2209.13266} [hep-ph].

\bibitem{Chang:2023cki}
C.~Chang, P.~Scott, T.~E. Gonzalo, F.~Kahlhoefer, and M.~White, Eur. Phys. J. C \textbf{83} (2023), no.~8, 692, \texttt{arXiv:2303.08351} [hep-ph], [Erratum: Eur.Phys.J.C 83, 768 (2023)].

\bibitem{GAMBIT:2021rlp}
GAMBIT, P.~Athron et~al., Eur. Phys. J. C \textbf{81} (2021), no.~11, 992, \texttt{arXiv:2106.02056} [hep-ph].

\bibitem{Planck:2018vyg}
Planck, N.~Aghanim et~al., Astron. Astrophys. \textbf{641} (2020), A6, \texttt{arXiv:1807.06209} [astro-ph.CO], [Erratum: Astron.Astrophys. 652, C4 (2021)].

\bibitem{Belanger:2011ww}
G.~Belanger and J.-C. Park, JCAP \textbf{03} (2012), 038, \texttt{arXiv:1112.4491} [hep-ph].

\bibitem{DAgnolo:2017dbv}
R.~T. D'Agnolo, D.~Pappadopulo, and J.~T. Ruderman, Phys. Rev. Lett. \textbf{119} (2017), no.~6, 061102, \texttt{arXiv:1705.08450} [hep-ph].

\bibitem{Garny:2017rxs}
M.~Garny, J.~Heisig, B.~L\"ulf, and S.~Vogl, Phys. Rev. D \textbf{96} (2017), no.~10, 103521, \texttt{arXiv:1705.09292} [hep-ph].

\bibitem{Brummer:2019inq}
F.~Br\"ummer, JHEP \textbf{01} (2020), 113, \texttt{arXiv:1910.01549} [hep-ph].

\bibitem{Kim:2019udq}
H.~Kim and E.~Kuflik, Phys. Rev. Lett. \textbf{123} (2019), no.~19, 191801, \texttt{arXiv:1906.00981} [hep-ph].

\bibitem{Dror:2016rxc}
J.~A. Dror, E.~Kuflik, and W.~H. Ng, Phys. Rev. Lett. \textbf{117} (2016), no.~21, 211801, \texttt{arXiv:1607.03110} [hep-ph].

\bibitem{Farina:2016llk}
M.~Farina, D.~Pappadopulo, J.~T. Ruderman, and G.~Trevisan, JHEP \textbf{12} (2016), 039, \texttt{arXiv:1607.03108} [hep-ph].

\bibitem{Puetter:2022ucx}
L.~Puetter, J.~T. Ruderman, E.~Salvioni, and B.~Shakya, Phys. Rev. D \textbf{109} (2024), no.~2, 023032, \texttt{arXiv:2208.08453} [hep-ph].

\bibitem{Maity:2019hre}
T.~N. Maity and T.~S. Ray, Phys. Rev. D \textbf{101} (2020), no.~10, 103013, \texttt{arXiv:1908.10343} [hep-ph].

\bibitem{Beauchesne:2024zsq}
H.~Beauchesne and C.-W. Chiang, Eur. Phys. J. C \textbf{84} (2024), no.~9, 950, \texttt{arXiv:2401.03657} [hep-ph].

\bibitem{Duan:2024urq}
X.-C. Duan, R.~Ramos, and Y.-L.~S. Tsai, Phys. Rev. D \textbf{110} (2024), no.~6, 063535, \texttt{arXiv:2404.12019} [hep-ph].

\bibitem{Cervantes:2024ipg}
E.~Cervantes and A.~Hryczuk, JHEP \textbf{11} (2024), 050, \texttt{arXiv:2407.12104} [hep-ph].

\bibitem{Boehm:2014hva}
C.~Boehm, M.~J. Dolan, C.~McCabe, M.~Spannowsky, and C.~J. Wallace, JCAP \textbf{05} (2014), 009, \texttt{arXiv:1401.6458} [hep-ph].

\bibitem{Hooper:2010mq}
D.~Hooper and L.~Goodenough, Phys. Lett. B \textbf{697} (2011), 412, \texttt{arXiv:1010.2752} [hep-ph].

\bibitem{Leane:2022bfm}
R.~K. Leane et~al., \texttt{arXiv:2203.06859} [hep-ph].

\bibitem{DAmbrosio:2002vsn}
G.~D'Ambrosio, G.~F. Giudice, G.~Isidori, and A.~Strumia, Nucl. Phys. B \textbf{645} (2002), 155, \texttt{hep-ph/0207036}.

\bibitem{Dolan:2014ska}
M.~J. Dolan, F.~Kahlhoefer, C.~McCabe, and K.~Schmidt-Hoberg, JHEP \textbf{03} (2015), 171, \texttt{arXiv:1412.5174} [hep-ph], [Erratum: JHEP 07, 103 (2015)].

\bibitem{Buckley:2014fba}
M.~R. Buckley, D.~Feld, and D.~Goncalves, Phys. Rev. D \textbf{91} (2015), 015017, \texttt{arXiv:1410.6497} [hep-ph].

\bibitem{DiMauro:2021qcf}
M.~Di~Mauro and M.~W. Winkler, Phys. Rev. D \textbf{103} (2021), no.~12, 123005, \texttt{arXiv:2101.11027} [astro-ph.HE].

\bibitem{Abdughani:2021oit}
M.~Abdughani, Y.-Z. Fan, C.-T. Lu, T.-P. Tang, and Y.-L.~S. Tsai, JHEP \textbf{07} (2022), 127, \texttt{arXiv:2111.02946} [astro-ph.HE].

\bibitem{Hryczuk:2022gay}
A.~Hryczuk and M.~Laletin, Phys. Rev. D \textbf{106} (2022), no.~2, 023007, \texttt{arXiv:2204.07078} [hep-ph].

\bibitem{Drees:2015exa}
M.~Drees, F.~Hajkarim, and E.~R. Schmitz, JCAP \textbf{06} (2015), 025, \texttt{arXiv:1503.03513} [hep-ph].

\bibitem{Bringmann:2006mu}
T.~Bringmann and S.~Hofmann, JCAP \textbf{04} (2007), 016, \texttt{hep-ph/0612238}, [Erratum: JCAP 03, E02 (2016)].

\bibitem{Binder:2016pnr}
T.~Binder, L.~Covi, A.~Kamada, H.~Murayama, T.~Takahashi, and N.~Yoshida, JCAP \textbf{11} (2016), 043, \texttt{arXiv:1602.07624} [hep-ph].

\bibitem{gondolo2012effect}
P.~Gondolo, J.~Hisano, and K.~Kadota, Physical Review D—Particles, Fields, Gravitation, and Cosmology \textbf{86} (2012), no.~8, 083523.

\bibitem{kasahara2009neutralino}
J.~Kasahara, \textit{Neutralino dark matter: the mass of the smallest halo and the golden region}, 2009.

\bibitem{Hannestad:2015tea}
S.~Hannestad, R.~S. Hansen, T.~Tram, and Y.~Y.~Y. Wong, JCAP \textbf{08} (2015), 019, \texttt{arXiv:1506.05266} [hep-ph].

\bibitem{Hahn-Woernle:2009jyb}
F.~Hahn-Woernle, M.~Plumacher, and Y.~Y.~Y. Wong, JCAP \textbf{08} (2009), 028, \texttt{arXiv:0907.0205} [hep-ph].

\bibitem{Cholis:2021rpp}
I.~Cholis, Y.-M. Zhong, S.~D. McDermott, and J.~P. Surdutovich, Phys. Rev. D \textbf{105} (2022), no.~10, 103023, \texttt{arXiv:2112.09706} [astro-ph.HE].

\bibitem{Cirelli:2010xx}
M.~Cirelli, G.~Corcella, A.~Hektor, G.~Hutsi, M.~Kadastik, P.~Panci, M.~Raidal, F.~Sala, and A.~Strumia, JCAP \textbf{03} (2011), 051, \texttt{arXiv:1012.4515} [hep-ph], [Erratum: JCAP 10, E01 (2012)].

\bibitem{vandenAarssen:2012ag}
L.~G. van~den Aarssen, T.~Bringmann, and Y.~C. Goedecke, Phys. Rev. D \textbf{85} (2012), 123512, \texttt{arXiv:1202.5456} [hep-ph].

\bibitem{Feng:2010zp}
J.~L. Feng, M.~Kaplinghat, and H.-B. Yu, Phys. Rev. D \textbf{82} (2010), 083525, \texttt{arXiv:1005.4678} [hep-ph].

\bibitem{DAgnolo:2015ujb}
R.~T. D'Agnolo and J.~T. Ruderman, Phys. Rev. Lett. \textbf{115} (2015), no.~6, 061301, \texttt{arXiv:1505.07107} [hep-ph].

\end{thebibliography}
\bibliographystyle{NewArXiv}
}
\end{document}